\documentclass[eqsecnum,nofootinbib,floats,preprint,aps,prd,floatfix,titlepage,tightenlines]
{revtex4} 
\usepackage{epsfig}
\usepackage{bm}

\makeatletter

\def\vereq#1#2{\lower3pt\vbox{\baselineskip1.5pt \lineskip1.5pt
\ialign{$\m@th#1\hfill##\hfil$\crcr#2\crcr\sim\crcr}}}

\@addtoreset{equation}{section} \makeatother

\newcommand{\bsigma}{{\mbox{\boldmath $\sigma$}}}
\newcommand{\btau}{{\mbox{\boldmath $\tau$}}}

\def\Tr{{\rm Tr\,}}
\newcommand{\sn}[2]{\,\mbox{sn}_#1\/#2}
\newcommand{\cn}[2]{\,\mbox{cn}_#1\/#2}
\newcommand{\dn}[2]{\,\mbox{dn}_#1\/#2}
\newcommand{\cosech}{\,\mbox{cosech}}

\newcommand{\high}{\vphantom{\Biggl(}\displaystyle}

\begin{document}

\title{Interactions of massless monopole clouds}

\author{Christopher M. Miller}
\email{cmm80@columbia.edu}
\author{Erick J. Weinberg}
\email{ejw@phys.columbia.edu}
\affiliation{Physics Department, Columbia University, New York, New York 10027
\vskip 0.5in}

\begin{abstract}

Some spontaneously broken gauge theories with unbroken non-Abelian
gauge groups contain massless magnetic monopoles that are realized
classically as clouds of non-Abelian field surrounding one or more
massive monopoles.  We use moduli space methods to investigate the
properties of these massless monopole clouds.  We show that the
natural metric on the Nahm data for a class of SU($2M+2$) solutions
with $2M$ massive and $M(2M-1)$ massless monopoles can be obtained
from that for a simpler class of SU($M+1$) solutions.  For the $M=1$
case, we show that the Nahm data metric is isomorphic to the metric
for the moduli of the BPS solutions, thus verifying a previously
conjectured result.  For $M=2$ we use our results and the moduli space
approximation to obtain an effective Lagrangian for an axially
symmetric class of solutions.  Using this Lagrangian, we study the
interactions between the two types of clouds that appear.  We show
that, although the static spacetime field configurations suggest that
the clouds might be rather diffuse, in scattering processes they
behave as if they were relatively thin hard shells.

\end{abstract}

\preprint{CU-TP-1187}

\maketitle

\section{Introduction}
\label{introduction}

$N=4$ supersymmetric Yang-Mills theory is believed to possess an
electric-magnetic duality symmetry.  When the gauge group is maximally
broken, to an Abelian subgroup, this duality interchanges the massive
electrically charged particles corresponding to the elementary fields
with magnetically charged states arising from classical soliton
solutions.  If the unbroken symmetry has a non-Abelian component, then
there exist massless particles carrying electric-type charge.
Although there are no isolated massless solitons, certain multisoliton
solutions have degrees of freedom that are naturally interpreted as
corresponding to massless magnetic monopoles.  These are manifested as
clouds of non-Abelian field that surround one or more massive
monopoles and shield part of their non-Abelian magnetic
charge~\cite{Lee:1996vz}.  Our aim in this paper will be to explore
the properties of these massless monopoles by studying the
interactions between these clouds.

To explain this in more detail, consider an SU($N$) gauge theory with an adjoint
Higgs field whose asymptotic value can be brought into the form
\begin{equation}
     \Phi  = {\rm diag}\, (s_1, s_2, \dots s_N)
\end{equation}
with $s_1 \le s_2 \le \dots s_N$.  If the $s_i$ are all distinct, the
gauge symmetry is broken maximally, to U(1)$^{N-1}$, and there are
$N-1$ topological charges.  If the asymptotic magnetic field is then written in
the form $F_{ij} = \epsilon_{ijk} r_k \, Q_M /r^3$, with 
\begin{equation}
     Q_M = {\rm diag}\, (n_1, n_2-n_1, \dots, -n_{N-1})  \, ,
\end{equation}
these topological charges are the $n_k$.  We will refer to an SU($N$) monopole
solution with such charges as being an $(n_1, n_2, \dots, n_{N-1})$
solution.  

With this maximal symmetry breaking, one can identify $N-1$
fundamental monopoles~\cite{Weinberg:1979zt}, each carrying a single
unit of one of the topological charges, with the mass of the $k$th
being\footnote{For the remainder of this paper, we will assume that the gauge
fields have been rescaled so as to set the
gauge
coupling $e$ to unity.}  $(4\pi/e)(s_{k+1}-s_k)$.  Each of these has
four degrees of freedom --- three position variables and one U(1)
phase.  A BPS solution with arbitrary magnetic charges can be
understood as being composed of appropriate numbers of the various
species of fundamental monopoles and as living on a moduli space whose
dimension is four times the total number of component monopoles.

Our interest here is not in maximal gauge symmetry breaking, but
rather in the alternative possibility, where $\Phi$ has degenerate
eigenvalues.  The unbroken symmetry is then enhanced to a non-Abelian
group, and some of the fundamental monopoles become massless.  It is
instructive to follow the behavior of the classical solutions as this
case is obtained from the maximally broken one by smoothly varying the
eigenvalues of $\Phi$.  One finds that an isolated fundamental
monopole solution goes over to the vacuum solution as it becomes
massless.  The behavior of multimonopole solutions is more
complex~\cite{Lu:1998br}.  If the total magnetic charge remains purely
Abelian, then the solution rapidly approaches its limiting form once
the inverse of the smallest monopole mass becomes larger than the
separations of the component monopoles.  In addition, the moduli space
of solutions and its naturally defined metric have smooth limits; in
the examples where the metric for the case with nonmaximal breaking
has been found directly, it is indeed the limit of the metric for the
maximally broken case~\cite{Lee:1996vz,Houghton:1999qu}.

If instead the magnetic charge has a non-Abelian component, none of
these are the case.  Not only do the moduli spaces not have smooth
limits, but cases that are gauge equivalent in the massless limit have
different dimensions for any finite monopole mass.  These pathologies
are clearly related to the long-range behavior of the non-Abelian
fields and the non-normalizability of certain zero modes in the
massless limit.  A simple example of this arises in an SU(3) gauge
theory.  With maximal breaking, the unbroken symmetry is
U(1)$\times$U(1), and there are two species of fundamental monopoles.
If two of the eigenvalues of $\Phi$ are equal, the unbroken symmetry
is SU(2)$\times$U(1) and one of the fundamental monopoles is massless.
In the maximally broken case, the moduli spaces of the (2,0) and the
(2,2) solutions have dimensions four and twelve, respectively.  Yet,
when the breaking is nonmaximal an SU(2) gauge transformation can turn
a (2,[0]) solution into a (2,[2]) solution, where the square brackets
denote the massless species.  On the other hand, the (2,[1])
solutions, which have a purely Abelian long-range field have a
well-defined moduli space, with a metric that is a smooth limit of the
(2,1) metric.

In this paper we will restrict ourselves to configuration with purely
Abelian magnetic charge.  Several such solutions with a single massless
monopole are known.  These include solutions~\cite{Weinberg:1982jh}
with one massive and one massless monopole for SO(5) broken to
SU(2)$\times$U(1), as well as SU(4) (1,[1],1)
solutions~\cite{Weinberg:1998hn} and the SU(3) (2,[1])
solutions~\cite{Dancer:1992kj} referred to above, each with one
massless and two massive monopoles.  These SU(3) solutions will be of
particular importance in our analysis.  Their properties were studied
in considerable detail by Dancer and
collaborators~\cite{Dancer:1992kj,Dancer:1992kn,Dancer:1992hf,Dancer:1997zx},
and we will refer to them, and their $(N-1,[N-2],\dots,[1])$
generalizations for SU($N$), as Dancer solutions.

In all of these solutions there is a single non-Abelian cloud
(spherical in the first case, ellipsoidal in the latter two) that
encloses the massive monopole(s).  Well inside this cloud, the
magnetic field approximates the field, with both Abelian and
non-Abelian magnetic components, that would be expected to arise just
from the massive monopoles.  The cloud effectively shields the
non-Abelian components, so that outside the cloud one sees the purely
Abelian field corresponding to the sum of the massive and massless
monopole charges.  The size of the cloud is
determined by a single parameter, whose value has no effect on the
total energy of the solution.

Since our goal in this paper is to investigate the interaction between
massless monopole clouds, we need solutions that have more than one
cloud.  One might have expected that these could be obtained simply by
having more than one massless monopole.  This turns out not to be so.
For example, the SU($N$) (1,[1],\dots,[1],1) solutions contain $N-3$
massless monopoles, but only a single cloud, no matter how large $N$
becomes~\cite{Lee:1996vz}.  [In fact, for $N>4$ these solutions are
essentially embeddings of (1,[1],1) SU(4) solutions into the larger
group.]

A set of solutions that do have multiple clouds, and which we will focus
on, are the (2,[2],[2],[2],2) solutions --- with four massive and six
massless monopoles --- in the theory with SU(6) broken to
U(1)$\times$SU(4)$\times$U(1).  The structure of these solutions was
studied in Ref.~\cite{Houghton:2002bz}.  They can be viewed as
containing two SU(3) Dancer solutions, each with a ``Dancer cloud''
enclosing two massive monopoles, embedded in disjoint subgroups of the
SU(6).  In addition, there are two larger ``SU(4) clouds'', enclosing
both Dancer clouds, that are somewhat analogous to the cloud in the
SU(4) (1,[1],1) solutions.  As we will describe in more detail later,
these four clouds are characterized not only by individual cloud-size
parameters, but also by a number of additional parameters specifying
their orientations with respect to the unbroken gauge group.  Note
that, even in this case, the number of clouds is less than the number
of massless monopoles.

A useful tool for studying the low-energy dynamics of multimonopole
systems such as these is the moduli space
approximation~\cite{Manton:1981mp}, which reduces the full field
theory dynamics to that of a finite number of collective coordinates
$z^a$.  The latter are governed by the Lagrangian
\begin{equation}
    L = {1\over 2} g_{ab} \dot z^a \dot z^b  \, ,
\end{equation}
where $g_{ab}$ is the metric on the moduli space of
BPS solutions.  This method was used to study the (1,[1],1) SU(4)
solutions~\cite{Chen:2001ge}.  Because the two massive monopoles lie
in mutually commuting subgroups of the SU($4$), one would not expect
them to interact directly with each other.  This turns out to be the
case; indeed, they can pass though each other undeflected.  On the
other hand, they do interact with, and exchange energy with, the
cloud.  In these interactions the cloud acts as if it were a thin
shell, with the monopole-cloud interactions concentrated in the times
when the massive monopole positions coincide with the shell.  At large
times the massive monopoles and the cloud decouple from each other and
evolve independently.  The interactions between the cloud and the
massless monopoles are similar in the SU(3) Dancer
solutions~\cite{Dancer:1992kn,Dancer:1992hf}, which have the additional
feature that the massive monopoles, being both of the same type, also
interact directly with each other.

A word of caution is in order here.  The essential idea underlying the
moduli space approximation is that for a slowly moving soliton
fluctuations off of the moduli space are energetically suppressed.  If
the theory contains massless particles, this needs further
examination, since excitation of these modes by radiation of massless
particles is always energetically possible.  However, since the source
of the radiation is proportional to the time derivative of the bosonic
fields, the radiation rate is expected to be small for low soliton
velocities.  This has been shown rigorously for configurations
involving pairs of monopoles in theories where the massless gauge
fields are all Abelian \cite{Manton:1988bn,Stuart:1994tc}.  Although
the validity of the approximation has not been demonstrated rigorously
for the case where there are massless non-Abelian gauge fields,
explicit comparison of the predictions of the moduli space
approximation with numerical evolution of the full field equations in
a spherically symmetric example~\cite{Chen:2001qt} show that
they agree as long as the configurations are slowly varying.

Of course, this method requires that moduli space metric be known.  It
can be obtained directly from the BPS solutions, if they are known
explicitly.  In some other cases indirect methods based on the
mathematical properties of the moduli space can be used to obtain
$g_{ab}$~\cite{Atiyah:1985dv,Lee:1996if,Gauntlett:1996cw}.  However,
for other cases it turns out to be easiest to resort to an alternative
approach.  The Atiyah-Drinfeld-Hitchin-Manin-Nahm (ADHMN)
construction~\cite{Nahm:1982jb,Nahm:1981xg,Nahm:1981nb,Nahm:1983sv} is
a powerful tool for obtaining BPS solutions.  It is based on an
equivalence between the Bogomolny equation for the fields in
three-dimensional space and an ordinary differential equation for a
set of matrix functions of a single variable, known as the Nahm data.
The moduli space of Nahm data has its own naturally defined metric.
It has been shown for both the SU(2) theory~\cite{nakajima} and for
the case of SU($N+1$) broken to U($N$)~\cite{takahasi}, and is
believed to be true in general, that the moduli spaces of Nahm data
and of BPS solutions are isometric.\footnote{In fact, we will
demonstrate this equivalence for yet another example, the (1,[1],1)
SU(4) solutions, in Sec.~\ref{oneoneone}.}  [In particular, Dancer
used this equivalence in his investigation of the dynamics of the
SU(3) solutions, and worked with the Nahm data metric.]  In this paper
we will assume that the equivalence of the two metrics holds as well
for our SU(6) example, and will work with the Nahm data metric.

The remainder of this paper is organized as follows.  In
Sec.~\ref{Nahmsec} we review the relevant parts of the ADHMN
construction, as well as the metric on the moduli space of Nahm data.
We then show that if the metric for the SU($M+1$) Dancer solutions is
known, then that for the $(M,[M],\dots,[M],M)$ solutions of SU($2M+2$)
broken to U(1)$\times$SU($2M$)$\times$U(1) can be easily obtained.  In
Sec.~\ref{oneoneone} we illustrate this method by obtaining the metric
for the (1,[1],1) SU(4) solutions and verify that the Nahm data metric
thus obtained is in fact equivalent to the metric on the moduli space
of BPS solutions, which had been found
previously~\cite{Lee:1996vz,Chen:2001ge,Lee:1996kz}.  Next, in
Sec.~\ref{twotwotwo}, we turn to the moduli space metric for the SU(6)
(2,[2],[2],[2],2) solutions.  These are described by a total of 40
collective coordinates.  Although the full metric can be obtained by
the methods of Sec.~\ref{Nahmsec}, the result would be rather unwieldy
for exploring the nature of the cloud dynamics.  We therefore consider
a restricted problem, that of a lower-dimensional submanifold of
axially symmetric solutions, for which the metric has a relatively
simple closed form expression but which still has enough structure to
allow us to investigate nontrivial cloud dynamics.  In
Sec.~\ref{dynamics} we use the metric obtained in the previous section
to explore the cloud-cloud interactions.  Section~\ref{conclude}
contains some concluding remarks.  There is an Appendix, which
contains some of the details of the moduli space metric calculation.

\section{The ADHMN construction and the metric for 
$\bm{(M,[M],\dots,[M],M)}$ solutions}
\label{Nahmsec}

As was discussed in Sec.~\ref{introduction}, the moduli spaces of BPS
solutions and of Nahm data are believed to be isometric.  We work
in this paper with the Nahm data metric, which is the more accessible
of the two for the examples that we study.  In this section we first
review the essential elements of the ADHMN
construction~\cite{Nahm:1982jb,Nahm:1981xg,Nahm:1981nb,Nahm:1983sv},
emphasizing the points that are relevant for the $(M,[M],\dots,M)$
solutions of SU($2M+2$).  We then obtain a general expression for the
metric of these SU($2M+2$) solutions, and briefly discuss an
asymptotic special case.

\subsection{Nahm data}

The basic elements in the ADHMN construction\footnote{For a fuller
description of the ADHMN construction, including a discussion of how
the spacetime fields are obtained from the Nahm data, see
Ref.~\cite{Weinberg:2006rq}.}  are the Nahm data, a quadruple of
matrix functions $T_\mu(s)$ ($\mu=0,1,2,3$) that obey the Nahm
equation,
\begin{equation}
   0=  {dT_i \over ds} + i[T_0,T_i] + {i\over 2} \epsilon_{ijk}[T_j,T_k]
     \, , \qquad i,j,k = 1,2,3   \, .
\label{nahmeq}
\end{equation}
For charge $k$ solutions in an SU(2) theory with Higgs vacuum
expectation value $v$, the $T_\mu(s)$ are $k\times k$ Hermitian matrices
defined for $-v/2 \le s \le v/2$.  For a general SU($N$)
theory, the eigenvalues of the Higgs vacuum expectation value divide
the range of $s$ into $N-1$ intervals, on each of which
Eq.~(\ref{nahmeq}) must hold.  (The boundary conditions at the ends of
these intervals are somewhat involved; we describe them below for the cases
we need.)
The dimension of the $T_\mu$ varies from 
interval to interval, being determined on each by the corresponding
magnetic charge.  If the $T_\mu$ are of the same size, $k\times k$, on
two adjacent intervals, then there are additional ``jumping data'',
forming a $2k$-component complex vector, associated with the boundary
between these intervals.

For later reference, note that if the $T_\mu(s)$ satisfy the Nahm
equation, then so do the $\tilde T_\mu$ defined by
\begin{eqnarray}
   \tilde T_i(s) &=& R_{ij} T_j(s) + B_i  \, , \cr
   \tilde T_0(s) &=& T_0(s)   \, ,
\label{rotANDtrans}
\end{eqnarray}
where $R_{ij}$ is an orthogonal matrix with determinant one.  The
$\tilde T_\mu(s)$ lead to a spacetime solution that is obtained 
from the original one by a combination of the spatial rotation specified by $R$
and a spatial translation by the vector $\bf B$.

For SU($2M+2$) broken to
U(1)$\times$SU($2M$)$\times$U(1) the eigenvalues of the Higgs vacuum expectation value
are $s_L < s_0 <s_R$, with $s_0$ being $2M$-fold degenerate.  
These divide the range of $s$ into a ``left'' interval
$[s_L,s_0]$, a ``right'' interval $[s_0,s_R]$, and $2M-1$ intervals of
zero width at $s=s_0$.  These correspond to two species of massive
monopoles, with masses
\begin{equation}
     M_L = 4\pi(s_0 - s_L) \, , \qquad M_R = 4\pi( s_R- s_0)  \, ,
\end{equation}
and $2M-1$ species of massless fundamental monopoles.  For the $(M,
[M], \dots, [M], M)$ solutions we need two sets of $M\times M$
matrices, $T_\mu^L(s)$ and $T_\mu^R(s)$, defined on the left and right
intervals, respectively.\footnote{The spacetime fields are obtained
  from sums of integrals over the various intervals, with the
  integrands obtained by solving a differential equation involving the
  Nahm matrices on the corresponding interval.  Because the integrals
  for the zero-width intervals vanish, the corresponding Nahm matrices
  have no effect on the spacetime fields.  For more details, see
  Ref.~\cite{Weinberg:2006rq}.}  These obey Eq.~(\ref{nahmeq}) subject
to the boundary condition that, for $M>1$, the $T_i^L$ ($T_i^R$) have
poles at $s_L$ ($s_R$) with the residues forming an $M$-dimensional
irreducible representation of SU(2).  Except for these poles, the
$T_\mu^L$ and $T_\mu^R$ must be everywhere nonsingular.

Because the magnetic charge is the same on adjacent intervals,
there are jump data associated with each of the $2M$ coincident
boundaries at $s_0$.  We write the data associated with the $F$th boundary
as a $2M$-component vector
$a_{\alpha r}^F$, where $r=1,2,\dots, M$ and $\alpha=1,2$.
These jump data are required to satisfy the constraint
\begin{equation}
   \left(\Delta T_j\right)_{rs}
      = \left[ T_j^L(s_0)\right]_{rs} - \left[T_j^R(s_0)\right]_{rs}
   = {1 \over 2}\sum_{F\alpha\beta}  a_{\alpha s}^{F*} 
          (\sigma_j)_{\alpha \beta} a_{\beta r}^F  \, .
\label{jumpeq}
\end{equation}
If we assemble the jump data into a $2M \times 2M$ matrix $A$ with
$A_{\alpha r, F} = a_{\alpha r}^F$,
and define an
$M\times M$ matrix
\begin{equation}
    (T_4)_{rs} = {1 \over 2}\sum_{F\alpha} a_{\alpha s}^{F*} a_{\alpha r}^F
\end{equation}
and a $2M\times 2M$ matrix
\begin{equation}
   K_{\alpha r; \beta s} \equiv (\Delta T_j)_{rs}  (\sigma_j)_{\alpha\beta}
         + (T_4)_{rs} \delta_{\alpha\beta}
        =   \sum_F a_{\beta s}^{F*} a_{\alpha r}^F  \, ,
\label{KmatrixDef}
\end{equation}
the jump data constraint becomes simply $K = AA^\dagger$.  For later
reference, it is important to note that this implies that the
eigenvalues of $K$ must be positive.  The general solution of this
constraint is
\begin{equation}
    A = K^{1/2} V \, ,
\label{Adecomp}
\end{equation}
where $V$ is unitary.  The freedom in the choice of $V$ reflects the
existence of an unbroken U(1)$\times$SU($2M$) subgroup of the original SU($2M+2$)
gauge group.\footnote{In the next subsection we will see how the effects of the 
remaining unbroken U(1) factor are manifested.}

The crucial observation for us is that the dimensions and boundary
conditions for the $T_\mu^L$ and $T_\mu^R$ are the same as those for
the Nahm data of the SU($M+1$) Dancer solution.  
Apart from the gauge orientation angles and phases encoded in $V$, 
the new information
specific to the SU($2M+2$) problem --- i.e., the parameters that
describe the SU($2M$) clouds --- enters only through $T_4$, which
can be chosen to be any Hermitian matrix, subject only to the
constraint that the eigenvalues of $K$ must be positive.  Thus, if the
Nahm equation has already been solved for the SU($M+1$) Dancer problem,
the only additional data needed for the $(M, [M], \dots, [M], M)$
solutions are the jump data, which are given by Eq.~(\ref{Adecomp}).

\subsection{Gauge action}

In addition to the spacetime symmetries described by Eq.~(\ref{rotANDtrans}), the
Nahm equations
have a set of invariances that are analogous to,
although distinct from, the gauge transformations on the spacetime
fields.  If $g(s)$ is a unitary matrix of appropriate dimension, the
Nahm equation (\ref{nahmeq}) and the jump equation (\ref{jumpeq}) are
invariant under\footnote{Such transformations are often used to make
$T_0(s)$ vanish identically.}
\begin{eqnarray}
     T_\mu(s) &\rightarrow& \tilde T_\mu(s) = g(s) T_\mu(s) g^{-1}(s)
      +  i \delta_{\mu 0} {dg\over ds} g^{-1}(s)  \, , \cr \cr
         A_{\alpha r, F} & \rightarrow& \tilde A_{\alpha r, F}
     = g(s_0)_{rs} A_{\alpha s, F}  \, .
\end{eqnarray}

When considering spacetime gauge transformations one distinguishes
between local gauge transformations, which approach the identity at
spatial infinity, and global gauge transformations, which are
nontrivial as $r \rightarrow \infty$.  While the former simply reflect
the presence of redundant field components, the latter correspond to
symmetries that are related to conserved gauge charges.  When the
gauge symmetry is unbroken, the number of global gauge transformations
is equal to the dimension of the gauge group.  If the symmetry is
spontaneously broken, as it must be when monopoles are present, only
those global gauge transformations that leave the asymptotic Higgs
field invariant (i.e., those in the unbroken gauge group) lead to
normalizable zero modes about a static solution, and only these
correspond to physical motions on the moduli space.

Similarly, we can distinguish between local gauge actions on the Nahm
data, for which $g(s)=I$ at both boundaries, and global gauge actions,
which have $g \ne I$ at one or both boundaries.  Any of the latter
that act on pole terms in the Nahm data will lead to nonnormalizable
modes, and therefore do not contribute to the moduli space dynamics.
This means that for the $(M, [M], \dots, M)$ solutions, which have
poles at both boundaries, the only relevant global gauge actions are
those that leave the pole terms invariant.
These are proportional to the
unit matrix and are of the form $g(s) = e^{i\chi(s)}I$.  Because an
$s$-independent gauge action proportional to the unit matrix would
have no effect on the Nahm data, it is sufficient to consider the case
where $\chi$ vanishes at one boundary, but not at the other; this
leads to a zero mode corresponding to an unbroken U(1).  The zero
modes corresponding to the other unbroken generators do not arise from
gauge actions, but instead correspond to variations of $V$.

For the SU($M+1$) Dancer solutions there must be a pole at one boundary, but the
only constraint at the other is that the Nahm data be nonsingular.
There are then a total of $M^2$ independent normalizable global gauge
zero modes, corresponding to the generators of the unbroken U($M$).
These can be obtained from gauge actions for which $g$ is an
$M$-dimensional unit matrix at the pole and proportional to one of
the generators of U($M$) at the other boundary.

\subsection{The moduli space metric}
\label{generalitiesSec}

The moduli space of Nahm data is the space of solutions of the Nahm and jump
equations, but with solutions that are related by local gauge actions
considered equivalent.  The coordinates on this space are the collective
coordinates $z_a$.  Its tangent space at a given point is spanned by
the variations of the Nahm data that preserve the Nahm and jump equations
and that are orthogonal to the variations
due to local gauge actions.

For our solutions, the Nahm data consist of two quadruples of Nahm
matrices, one for each interval, and jump data at the boundary at $s_0$.
We can write these collectively as ${\cal T}=\{T^L_\mu(s),
T^R_\mu(s), A\}$.
The tangent vectors to the Nahm data moduli space must have a similar
structure, and so can be written as ${\cal Y} = \{ Y^L_\mu(s),
Y^R_\mu(s), Y\}$.  The inner product of two such vectors is defined to
be\footnote{Note that in the trace in the last term the
indices run over $2M$ values, whereas in the first two terms the
traces are of $M\times M$ matrices.}
\begin{equation}
    \langle {\cal Y}, {\cal Y'} \rangle
        =  \int_{s_L}^{s_0} \, \Tr Y^L_\mu(s) Y^{'L}_\mu(s)
        + \int_{s_0}^{s_R} \, \Tr Y^R_\mu(s) Y^{'R}_\mu(s)
        +  {1\over 2}\Tr (Y Y^{'\dagger} + Y' Y^\dagger )  \, .
\end{equation}

An infinitesimal local gauge action is specified by a Hermitian matrix
function $\Lambda(s)$ that is everywhere continuous and vanishes at
$s_L$ and $s_R$.  We denote its values on the left and right intervals
by $\Lambda^L(s)$ and $\Lambda^R(s)$, and define $\Lambda(s_0) \equiv
\Lambda^0$.  Its action on the Nahm data defines a vector ${\cal
Y}_{\rm gauge}$ with  
\pagebreak
\begin{eqnarray}
   (Y_{\rm gauge})_\mu^{L,R} &=&  \delta_\Lambda T_\mu^{L,R} 
         = \delta_{\mu 0} {d\Lambda^{L,R} \over
          ds} +i[T_\mu^{L,R}, \Lambda^{L,R}] \equiv D_\mu \Lambda^{L,R}
        \, , \cr \cr 
   (Y_{\rm gauge})_{\alpha r, F} &=& (\delta_\Lambda A)_{\alpha r, F}
     = -i \Lambda^0_{rs} A_{\alpha s, F}  \, ,
\end{eqnarray}
where $D_j = i[T_j(s),~]$ and $D_0= d/ds + i[T_0(s),~]$.  In order
that $\cal Y$ be orthogonal to ${\cal Y}_{\rm gauge}$ for any choice
of $\Lambda$, we must require that
\begin{eqnarray}
     0 &=& D_\mu Y_\mu^{L,R}(s) \, , \qquad s\ne s_0  \, ,
\label{backgroundgauge}   \\
     0 &=& \left[Y_0^L(s_0) - Y_0^R(s_0)\right]_{rs}
     + {i\over 2} (YA^\dagger - AY^\dagger)_{\alpha r, \alpha s}  \, . 
\label{jumpgaugecondition}
\end{eqnarray}
By analogy with the corresponding constraint on the variations of the
spacetime fields, we will refer to these as the background gauge
conditions.

A basis for the tangent space is given by a set of vectors ${\cal Y}_a$
of the form
\begin{eqnarray}
     Y_{a\mu}^{L,R}(s) &=& {\partial T_\mu^{L,R}(s;z) \over \partial z^a}
            + D_\mu \Lambda_a^{L,R}  \, , \cr\cr
  Y_a &=& {\partial A \over \partial z^a} - i[\Lambda_a(s_0) \otimes A] \, ,
\label{YisTplusgauge}
\end{eqnarray}
with the gauge action
$\Lambda_a(s)$ chosen so that ${\cal Y}_a$ is in background gauge.
The metric on the moduli space is then defined to
be\footnote{The factor of $4\pi$ here is chosen to make the
normalizations of the Nahm data metric and the BPS solution
metric the same; it can be easily checked by noting the coefficient of
the terms quadratic in the center-of-mass position.}

\begin{equation}
     ds^2 =  g_{ab}\,  dz^a dz^b
      = 4 \pi \langle {\cal Y}_a, {\cal Y}_b \rangle \,  dz^a dz^b  \, .
\label{metricdef}
\end{equation}

\subsection{The metric for the $\bm {(M, [M], \dots, M)}$ solutions of SU($\bm {2M+2}$)}
\label{Nahmmetric}

As we have just seen, to calculate the moduli space metric one must
first vary the Nahm data with respect to the coordinates, and then
find a gauge action $\Lambda(s)$ that brings the resulting tangent
vector into background gauge.  The one part of this procedure that is
not completely straightforward is the solution of the differential
equation for $\Lambda(s)$ that is implied by
Eqs.~(\ref{backgroundgauge}) and (\ref{YisTplusgauge}).  However, if
this has already been done for the SU($M+1$) Dancer solutions, the
determination of the moduli space metric for the $(M, [M], \dots, M)$
solutions of SU($2M+2$) reduces to an algebraic problem.  To see this,
first note that the Dancer problems give us two sets of basis vectors,
$Y^{DL}_{a\mu}$ and $Y^{DR}_{a\mu}$, that satisfy
Eq.~(\ref{backgroundgauge}) on their respective domains.  Each of
these sets includes $M^2$ vectors corresponding to the global U($M$)
gauge freedom.  These are of the form $Y^{L}_{f\mu}=D_\mu\chi_f^L$ and
$Y^{R}_{f\mu}=D_\mu\chi^R_f$ ($f=1,2,\dots M^2$), where $\chi_f^L$
($\chi_f^R$) vanishes at $s_L$ ($s_R$) and is nonzero and proportional
to one of the U($M$) generators at $s_0$.  Because the Dancer basis
vectors satisfy Eq.~(\ref{backgroundgauge}),
\begin{equation}
     D_\mu D_\mu \chi_f^L =  D_\mu D_\mu \chi_f^R = 0  \, .
\label{d2Lambda}
\end{equation}

Now suppose that the coordinates for the $(M, [M], \dots, M)$ solutions
are chosen so that 
one subset are those originating
with the left Dancer problem, a second subset are those from the right
Dancer problem, and the remainder are associated only with the jump
data.  The tangent vector corresponding to the coordinate $z^a$ can then be
written in the form
\begin{equation}
   {\cal Y}_a = \left\{ Y_{a\mu}^{DL} +  D_\mu \Lambda^L_a, 
             Y_{a\mu}^{DR} +  D_\mu \Lambda^R_a,
    {\partial A \over \partial z^a}  -i(\Lambda_a^0 \otimes I_2)A\right\} \, ,
\end{equation}
where $Y_{a\mu}^{DL}$ and $Y_{a\mu}^{DR}$ are the vectors from the
left and right Dancer problems.\footnote{Of course, with coordinates
chosen as described above, at most one of these Dancer vectors will be
nonzero for a given $z^a$.}  Although these Dancer vectors already
satisfy the background gauge condition on their respective domains, an
additional gauge action may be needed to satisfy
Eq.~(\ref{jumpgaugecondition}) at $s_0$.  Its gauge function $\Lambda$
must vanish at $s_L$ and $s_R$ and, in order to maintain the
background gauge condition on the two intervals, it must satisfy
$D_\mu D_\mu \Lambda =0$ for $s \ne s_0$.  It is therefore a
linear combination of Dancer global gauge modes, with
\begin{eqnarray}
      \Lambda^L_a &=& c^L_{af} \chi_f^L \, , \cr
      \Lambda^R_a &=& c^R_{af} \chi_f^R  \, .
\label{gaugemodeexpansion}
\end{eqnarray}
Thus, for each ${\cal Y}_a$ there are
a total of $2M^2$ constants to be determined.
Requiring continuity of the gauge action at the boundary, $\Lambda^0_a=
\Lambda^L_a(s_0)= \Lambda^R_a(s_0)$, gives 
$M^2$ algebraic equations.  The
background gauge condition at the boundary becomes
\begin{eqnarray}
    \left[ D_0\Lambda^L_a(s_0) - D_0\Lambda^R_a(s_0) + \Lambda^0_a T_4
          + T_4 \Lambda^0_a \right]_{rs} &=& \left[ Y_{a0}^{DR}(s_0) -
          Y_{a0}^{DL}(s_0) \right]_{rs} 
       \cr\cr && \qquad
          +{i \over 2} \left(A
          {\partial A^\dagger \over \partial z_a} - {\partial A \over
          \partial z_a} A^\dagger\right)_{\alpha r,\alpha s}
\label{jumpcondition}
\end{eqnarray}
and gives $M^2$ more equations, thus determining the $c^L_{af}$ and
$c^R_{af}$, and hence ${\cal Y}_a$.

We can now evaluate the metric.  Because the ${\cal Y}_a$ satisfy the
background gauge conditions of Eqs.~(\ref{backgroundgauge}) and
(\ref{jumpgaugecondition}) and, in addition, $D_\mu Y_{a\mu}^{DL} =
D_\mu Y_{a\mu}^{DR} =0$, many of the terms involving the gauge actions
can be eliminated by integrations by parts.  With the aid of
Eq.~(\ref{jumpcondition}), one eventually obtains
\begin{eqnarray}
    g_{ab} &=& 4\pi \int_{s_L}^{s_0} ds \, \Tr Y_{a\mu}^{DL} Y_{b\mu}^{DL}
            + 4\pi \int_{s_0}^{s_R} ds \, \Tr Y_{a\mu}^{DR} Y_{b\mu}^{DR}
          + g_{ab}^0  \cr \cr
      &=& g_{ab}^{DL} + g_{ab}^{DR} +  g_{ab}^0  \, ,
\label{metricform}
\end{eqnarray}
where $g_{ab}^{DL}$ and $g_{ab}^{DR}$ are the metric
components from the corresponding Dancer solutions and
\begin{eqnarray}
     g_{ab}^0 &=& 4\pi \, 
     \Tr \left[Y_{a0}^{DL}(s_0) - Y_{a0}^{DR}(s_0)\right] \Lambda^0_b 
      + 2\pi \, \Tr \left({\partial A\over \partial z^a}        
                        {\partial A^\dagger\over \partial z^b}
      + {\partial A\over \partial z^b} {\partial A^\dagger\over \partial z^a} \right)
        \cr  &&\qquad
  + 2 \pi i \, \Tr \left[
          {\partial A\over \partial z^a} A^\dagger (\Lambda_b^0 \otimes I_2)
        - (\Lambda_b^0 \otimes I_2) A {\partial A^\dagger\over \partial z^a} \right]
   \cr\cr   &=& 2 \pi \, \Tr \left({\partial A\over \partial z^a}
                        {\partial A^\dagger\over \partial z^b} 
    + {\partial A\over \partial z^b} {\partial A^\dagger\over \partial z^a} \right)
     +  4\pi\, \Tr \left[D_0\Lambda^R_a(s_0) - D_0\Lambda^L_a(s_0)\right] \Lambda_b^0
       \cr  &&\qquad
     - 4\pi \,\Tr  T_4 \left(\Lambda_a^0 \Lambda_b^0 + \Lambda_b^0 \Lambda_a^0 \right)
\label{jumpmetric}
\end{eqnarray}
contains the entire contribution from the jump data.\footnote{Although
the middle term in the final expression for $g_{ab}^0$ appears not to
be symmetric under interchange of $a$ and $b$, it actually is.  This
can be shown by an integration by parts and making use of the facts
that the $\Lambda_c$ obey $D_\mu D_\mu \Lambda_c =0$ and vanish at
$s_L$ and $s_R$.}   These two equations are the main result of this 
section.

\subsection{Large SU($2M$) clouds}
\label{largeCloudSec}

Before focusing on the specific examples of $M=1$ and $M=2$,
it is worth commenting briefly on the case where the length scales in
$T_\mu^{DL}$ and $T_\mu^{DR}$ are smaller than all those entering
$T_4$ by a factor of $\epsilon \ll 1$. This corresponds to the
situation where the SU($2M$) clouds are large compared to both the
Dancer clouds and the separations between the massive monopoles.  In
this case Eq.~(\ref{Adecomp}) can be written as
\begin{equation}
    A = (T_4^{1/2}\otimes I_2)V +\delta A  \, , 
\end{equation} 
where $\delta A$, which contains all of the information about the
Dancer data, is suppressed by a factor of $\epsilon$.  Therefore, if
$z_a$ is one of the Dancer coordinates, $\partial A/\partial z_a =
O(\epsilon)$.  Furthermore, by noting the $\Lambda_a^0 T_4$ terms on
the left-hand side of Eq.~(\ref{jumpcondition}), we see that the
$\Lambda_a$ corresponding to these coordinates are also suppressed by a
factor of $\epsilon$.  It follows from these facts that (1) if either $a$ or
$b$ refers to a Dancer coordinate, then $g_{ab}^0$ is suppressed
relative to $g_{ab}^{DL}$ or $g_{ab}^{DR}$ and (2) if $a$ and $b$
both refer to jump coordinates, then all of the dependence of $g_{ab}^0$ on
Dancer parameters is through subleading terms.  Hence, to leading
order in $\epsilon$ the moduli space Lagrangian separates into two
parts, one depending only on the Dancer parameters and one
depending only on the jump parameters.  In other words, large SU($2M$)
clouds are effectively decoupled from both the massive monopoles and the
Dancer clouds.

\section{(1,[1],1) solutions in SU(4)}
\label{oneoneone}

We will first consider the (1,[1],1) solutions for a theory with SU(4)
broken to U(1)$\times$SU(2)$\times$U(1).  This will not only serve as
an illustration of our method but, because the metric on the
space of BPS solutions is already
known~\cite{Lee:1996vz,Chen:2001ge,Lee:1996kz}, it will also provide
one more example supporting the conjecture that the moduli spaces of
Nahm data and of BPS solutions are isometric.

The ``Dancer'' solutions in this case are just embeddings of the unit
SU(2) monopole, each with a four-dimensional moduli space.  The Nahm data
are numbers rather than matrices and are given (with a
standard choice of the gauge action) on the left interval 
by $T^L_0 =0$, $T^L_j = - X^L_j$, where
the $X^L_j$ are the coordinates of the monopole center.  Differentiating
these with respect to the $X^L_j$ gives three Dancer tangent vectors
\begin{equation}
     \left[Y_{X^L_j}^{DL}\right]_\mu = - \delta_{\mu j}
\end{equation}
that satisfy the background gauge condition without needing any
compensating gauge action.  The fourth tangent vector corresponds to a U(1)
phase, and so must be of the form 
\begin{equation}
     \left[Y_{\rm U(1)}^{DL}\right]_\mu = D_\mu \chi^L 
           = \delta_{\mu 0} {d\chi^L \over ds}   \, .
\end{equation}
In order that this be in background gauge, we need
that $d^2 \Lambda/ds^2 =0$.  Fixing the normalization by requiring
that $\chi^L(s_0) =1$ and  $\chi^L(s_L)=0$, we find that 
\begin{equation}
    \chi^L(s) = {4\pi(s-s_L) \over M_L} \, , \qquad   \left[Y_{\rm U(1)}^{DL}\right]_\mu = 
             \delta_{\mu 0}  \left( {4\pi \over M_L} \right)   \, .
\end{equation}

The Nahm data and tangent vectors for the right Dancer data are completely
analogous.  They are obtained simply by replacing $L$ by $R$ in the above
equations, except for a sign change that yields
\begin{equation}
    \chi^R(s) = -{4\pi(s-s_R) \over M_R} \, , \qquad   \left[Y_{\rm U(1)}^{DR}\right]_\mu =
            - \delta_{\mu 0}  \left( {4\pi \over M_R} \right)  \, .
\end{equation}
In the absence of jump data, these two sets of four vectors would give a
moduli space metric
\begin{equation}
   ds_L^2 + ds_R^2 = M_L \,d{\bf X}_L^2 +  M_R \,d{\bf X}_R^2 + 
        {(4\pi)^2\over M_L} d\chi_L^2 + {(4\pi)^2\over M_R} d\chi_R^2  \, .
\end{equation}

In the standard fashion, we can rewrite the positions in terms of
center-of-mass and relative positions $\bf X_{\rm CM}$ and $\bf R$.
We can also replace $\chi_L$ and $\chi_R$ by a global U(1) phase and a
relative U(1) phase.  The former, given by $\xi = \chi_L +
\chi_R$, corresponds to a simultaneous phase rotation of the two
monopoles, and is described by a tangent vector
$\{\left[Y^{DL}_\xi\right]_\mu, \left[Y^{DR}_\xi\right]_\mu\} =
4\pi\delta_{\mu 0}/(M_L+M_R) \{1,1\} = 4\pi/(M_L+M_R) \, D_\mu
\{M_L\,\chi^L, -M_R\,\chi^R\}$.  Note that although
$\left[Y^{DL}_\xi\right]_\mu$ and $\left[Y^{DR}_\xi\right]_\mu$
correspond to pure gauge actions on their separate intervals, the
combined vector is not a gauge action because the corresponding left
and right gauge functions are not equal at $s=s_0$.

The relative U(1) phase $\psi= (M_L\chi_R-M_R\chi_L)/(M_L+M_R) $ 
corresponds to an orthogonal combination of
vectors and, in the context of just the Dancer data, is a pure gauge
action.  We can therefore choose the gauge so that $\left[Y^{DL}_\psi\right]_\mu$ and 
$\left[Y^{DR}_{\psi}\right]_\mu$ both vanish and the $g_{a\psi}$ are given completely
by the jump data term $g^0_{a\psi}$

We now have to consider the contributions from the jump data.  We start
by defining $T_4 = b$; examination of the 
spacetime solutions shows that $b$ measures the size of the non-Abelian 
cloud.  We then have
\begin{equation}
       K = b I_2 + {\bf R}\cdot {\bsigma} = U K_0 U^{-1}  \, ,
\end{equation}
where $K_0 = {\rm diag}\, (b+R, b-R)$, and can write the general
solution for the jump data as
\begin{equation}
   A = U K_0^{1/2} W e^{i\psi}  \, ,
\end{equation}
where $W$, like $U$, is an SU(2) matrix.\footnote{In the notation of
Eq.~(\ref{Adecomp}), $W=U^{-1}V$.  We have written $A$ in this form to
facilitate comparison with the results in Ref.~\cite{Chen:2001ge}.}
Neither the center-of-mass position nor the global U(1) phase
$\xi_{\rm total}$ appears in $A$, so the tangent vectors corresponding
to these variables have no jump component and are specified completely
by the $Y_\mu^{DL}$ and $Y_\mu^{DR}$ inherited from the Dancer
problems.  It is easily verified that these vectors are in background
gauge and that they are orthogonal to the vectors for the relative
coordinates.

The calculation of the remaining metric terms is simplified by noting
that, because the spatial rotations represented by $U$ and the global
U(2) symmetry represented by $W$ and $\psi$ are isometries, we
can calculate the metric from the tangent vectors at a point with
$U=W=I$, $\psi=0$.  At this point the tangent vector for the intermonopole
separation $R$ gets a jump data contribution
\begin{equation}
     {\partial A\over \partial R} = {1 \over 2}  
       \,{\rm diag}\, \left({1\over \sqrt{b+R}}, -{1\over \sqrt{b-R}} \right)
\end{equation}
that combines with the contributions from the Dancer vectors to give 
a tangent vector ${\cal Y}_R$ that is already in background gauge.  For the cloud
size parameter $b$ we have
\begin{equation}
     {\partial A\over \partial b} = {1 \over 2}  
    \,{\rm diag}\, \left({1\over \sqrt{b+R}}, {1\over \sqrt{b-R}} \right) \, .
\end{equation}
Because $b$ does not enter the Dancer solutions, ${\cal Y}_b$ has no
$Y_{\mu}^{DL}$ or $Y_{\mu}^{DR}$ contribution.  It, too, satisfies the
background gauge conditions without the need for a compensating gauge
action.

To obtain the remaining tangent vectors, which correspond to rotations
and U(2) transformations, we first write an infinitesimal variation of
$A$ (with $K_0$ held fixed) as
\begin{equation}
    dA = {i\over 2} \sum_{i=1}^2  \sigma_j K_0^{1/2} \, d\alpha_j 
        + {i\over 2} \sum_{i=1}^3 K_0^{1/2} \sigma_j  \, d\beta_j
        + {i\over 2} K_0^{1/2} \, d\psi  \, ,
\end{equation}
where the $d\alpha_j$ and $d\beta_j$ are the invariant one-forms for
the rotational SO(3) and gauge SU(2) symmetries.\footnote{We have not
included an $\alpha_3$ term because its effect can be absorbed by a
redefinition of $\beta_3$; this term would correspond to the Euler
angle that leaves $\bf R$ invariant.}  The $d\alpha_1$ and $d\alpha_2$
terms combine with contributions, due to rotations of $\bf R$, from
the left and right intervals to give background gauge tangent vectors.
The $\beta_1$ and $\beta_2$ vectors have no contributions from these
intervals, but are also in background gauge.  However, the $\beta_3$
and $\psi$ vectors do not satisfy Eq.~(\ref{jumpgaugecondition}) and
so must be supplemented by compensating gauge actions.  As explained
in Sec.~\ref{Nahmmetric}, these gauge actions must have gauge
functions of the form $\Lambda_a^L = c^L_a \chi^L$, $\Lambda_a^R =
c^R_a \chi^R$.  Continuity of $\Lambda$ at $s_0$ implies that
$c^L_a=c^R_a\equiv c_a$.  Using Eq.~(\ref{jumpcondition}), we then
find that
\begin{equation}
      c_{\beta_3} = {R\mu \over 4\pi +2b\mu} \, ,   \qquad 
      c_{\psi} = {b\mu \over 4\pi +2b\mu}   \, ,
\end{equation}
where $\mu = M_L M_R /(M_L+M_R)$ is the reduced mass.

With the $\Lambda_a$ thus determined, we can use Eqs.~(\ref{metricform})
and (\ref{jumpmetric}) to 
show that the metric for the eight-dimensional relative
moduli space [i.e., with the center-of-mass motion and overall U(1)
phase factored out] is
\begin{eqnarray}
     ds^2 &=& \left[ \mu + {2\pi b \over (b^2 - R^2)} \right] dR^2
      + {2\pi b \over (b^2 - R^2)} db^2 - {4\pi R \over b^2 - R^2}\, db \, dR
       \cr\cr &&\quad
        + \left( \mu R^2 + {2\pi b}\right) (d\alpha_1^2 + d\alpha_2^3)
         + {2\pi b} (d\beta_1^2 + d\beta_2^2)
         + 4\pi \sqrt{b^2-R^2} (d\alpha_1 \, d\beta_1 + d\alpha_2\, d\beta_2)
       \cr\cr &&\quad
         + {4\pi^2 b \over 2\pi + b\mu }d\psi^2
    + \left[{2\pi b} - {2\pi \mu R^2 \over (2\pi + b\mu) }\right] d\beta_3^2
         + {8\pi^2 R\over (2\pi + b\mu) }\, d\psi \, d\beta_3   \, .
\end{eqnarray}
\pagebreak
This agrees with the metric on the moduli space of BPS solutions that
was previously obtained.\footnote{This is verified most easily by
comparing with the form given in Ref.~\cite{Chen:2001ge}.  For the
angular and phase parts of the metric our expression, written in terms of
angular velocities, is related to the one in Eq.~(3.18) of that
paper, given in terms of angular momenta, by a Legendre
transformation.}  This thus provides another example where the moduli
spaces of Nahm data and of BPS solutions are isometric, lending
further support to the conjecture that this is true in general.

For later reference, we note that when the angular momenta and charges all
vanish, the system reduces to one governed by the Lagrangian
\begin{equation}
    L = {\pi\over 2} {(\dot b + \dot R)^2\over (b+R)}
    + {\pi\over 2} {(\dot b - \dot R)^2\over (b-R)} 
    + {\mu \over 2} \dot R^2   \, .
\label{su4Lag}
\end{equation}

\section{(2,[2],[2],[2],2) solutions in SU(6)}
\label{twotwotwo}

\subsection{Nahm data}

The Nahm data are $2 \times 2$ matrix functions $T_\mu^L(s)$ and
$T_\mu^R(s)$ on the left and right intervals, respectively, plus jump data
that are obtained
from $T_\mu^L(s_0)$, 
$T_\mu^R(s_0)$, and 
\begin{equation}
    T_4 = p I_2 + {\bf q} \cdot \btau
\label{pqdef}
\end{equation}
by using Eqs.~(\ref{KmatrixDef}) and (\ref{Adecomp}).  The parameters in
$T_4$ determine the properties of the two SU(4) clouds.  Examination of the 
spacetime solutions~\cite{Houghton:2002bz} shows that,
roughly
speaking, $p+q$ and $p-q$ (with $q=|{\bf q}|$) determine the sizes of
the clouds, while the direction of $\bf q$ specifies an orientation in
the unbroken SU(4).  The $T_\mu^L(s)$ are themselves the Nahm data for
an SU(3) (2,[1]) Dancer solution.  By an appropriate gauge action one
can set $T_0^L=0$ and then write~\cite{Dancer:1992kn}
\begin{equation}
   T_i^L = {1\over 2} \sum_j A^L_{ij} f_j^L(s) \hat\tau_j^L  + R_i^L {\rm I}_2  \, ,
\label{dancermetricform}
\end{equation}
where $A_{ij}^L$ is an orthogonal matrix and the three $\hat\tau_j^L =
U_L \tau_j U_L^{-1}$ are a rotated set of Pauli matrices.  The
$f_j^L(s)$ obey
\begin{equation}
     {df_1^L \over ds} = f_2^L f_3^L
\label{topeq}
\end{equation}
and its two cyclic permutations.  If we adopt the convention that 
$f_1^2 \le f_2^2 \le f_3^2$, they  
are given in terms of 
Jacobi elliptic functions by
\begin{eqnarray}
    f_1^L(s) &=&  - {D_L \cn {{\kappa_L}} [D_L(s-s_L)] 
            \over \sn {{\kappa_L}} [D_L(s-s_L)] } \, , 
    \cr \cr
    f_2^L(s) &=&  - {D_L \dn {{\kappa_L}} [D_L(s-s_L)] 
            \over \sn {{\kappa_L}} [D_L(s-s_L)] }  \, , 
    \cr \cr
    f_3^L(s) &=&  - {D_L \over \sn {{\kappa_L}} [D_L(s-s_L)] }  \, .
\label{topfunctions}
\end{eqnarray}
The requirement that $f_j^L(s)$ only have a pole at $s_L$ imposes the 
conditions
$0\le \kappa_L \le 1$ and $0 \le D_L \le 2K(\kappa_L)/(s_0-s_L)$, 
where $K(\kappa)$ is the complete elliptic integral of the first kind. 
The $T_\mu^R(s)$ are similar, but with $D_L$ and $\kappa_L$ replaced by 
$D_R$ and $\kappa_R$.

\begin{figure}
\begin{center}
\leavevmode
\epsfysize=3in
\epsffile{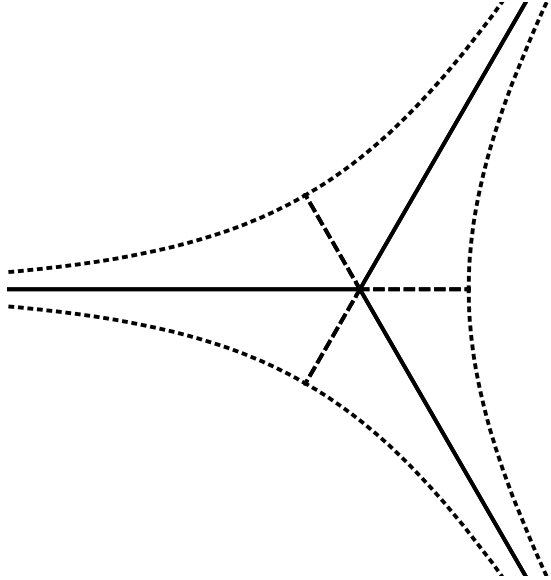}
\end{center}
\caption{A geodesically complete submanifold illustrating the SU(3)
Dancer solutions.  The long straight lines correspond to the axially
symmetric hyperbolic solutions, with two widely separated monopoles,
while the short straight lines correspond to the axially symmetry
trigonometric solutions.  The limiting curved boundaries, which are
not part of the manifold, correspond to SU(2) two-monopole solutions.}
\label{dancerspace}
\end{figure}

The left and right Nahm data each contain eleven parameters: three
center-of-mass variables $R_j$, the three Euler angles in $A_{ij}$
that specify the spatial orientation, the three angles needed to 
define the $\hat\tau_j$, and the elliptic function
parameters $D$ and $\kappa$.  The significance of the latter two is
clarified by referring to the plot in Fig.~\ref{dancerspace}.  The
change of variables~\cite{Dancer:1992hf}
\begin{eqnarray}
    x &=& (2 - \kappa^2) D^2 \, ,   \cr
    y &=& -\sqrt{3}\, \kappa^2 D^2
\end{eqnarray}
maps the allowed range of $D$ and $\kappa$ onto the lower right
sextant of the plot (including the straight boundaries, but excluding
the curved outer boundary, which is geodesically infinitely far from
any point in the interior).  By adjoining five other copies (corresponding to
the other possible orderings of the $f_j^2$), one
obtains a geodesically complete two-dimensional manifold.  Points far
out on the long arms of the figure correspond to solutions with two
well-separated massive monopoles, with the distance between the two
approximately equal to $D$.  The straight lines down the centers of the
arms, on which $\kappa =1$, correspond to minimal Dancer cloud size,
while the limiting curve corresponds to embeddings of SU(2)
two-monopole solutions that can be thought of as having infinite
Dancer clouds.  The central point, where $D=0$ and $\kappa$ is
undefined, corresponds to a solution with coincident massive monopoles
and a minimal size cloud.  On the short straight lines emanating from
this point $\kappa = 0$. Points on these lines correspond to solutions
with coincident massive monopoles and clouds varying from minimal to
infinite size~\cite{Dancer:1997zx,Irwin:1997ew}.

For $\kappa$ equal to 0 or 1, the elliptic functions reduce to
trigonometric or hyperbolic functions,
respectively~\cite{Dancer:1992kj}.  Two of the $f_j$ are then equal
and the spacetime solution has an axial symmetry.  When $D=0$, all
three of the $f_j$ are equal and the solution is spherically
symmetric.  A straight trajectory passing from a $\kappa =0$ line
through the central point and out along the opposite $\kappa = 1$ line
is a geodesic of the Dancer metric.

\subsection{Reduction to cylindrical symmetry with vanishing charges}
\label{cylindrical}

As noted above,  
the left and right sets of Dancer data each contain eleven parameters.
In addition, there is an overall U(1) phase associated with each set.
These, plus the four parameters from the elements
of $T_4$ given in Eq.~(\ref{pqdef}) and the sixteen moduli arising
from the U(4) matrix $V$ would seem to give a total of 44 moduli.
This cannot be correct, because a solution with ten monopoles should
lie on a 40-dimensional moduli space.  The discrepancy is resolved by
noting that there is a U(2) subgroup of the U(4) whose effect is gauge
equivalent to that obtained by simultaneously rotating the U(1) phases
and SU(2) orientations of the two Dancer solutions and the SU(2)
orientation of the vector $\bf q$.

Because the moduli space metric for the Dancer data is already known,
the methods of Sec.~\ref{Nahmsec} can be applied to obtain the metric
for the full 40-dimensional moduli space.  However, the result would
be rather unwieldy for exploring the nature of the cloud dynamics.  We
will therefore reduce the problem to a more manageable one by
restricting ourselves to a considerably smaller, but geodesically
complete, submanifold.  

A geodesically complete submanifold can be obtained by restricting to
the maximal subspace left invariant by some isometry of the full
manifold.  In particular, we will require that the solutions be
axially symmetric about the $z$-axis.  This means that the Nahm data
must be invariant under the combination of a rotational transformation
of the form given in Eq.~(\ref{rotANDtrans}) and an appropriately
chosen gauge action.  In each set of
Dancer data two of the $f_j$ must then be equal
(which is only possible if $\kappa=0$ or 1), which implies that
the solution acts as a symmetric top in the SU(2) space.  Furthermore,
the symmetry axes of the two Dancer solutions must be aligned with
each other and with $\bf q$.  More specifically, the $\hat \tau_j^L$ and 
the $\hat \tau_j^R$ can differ only by an U(1) rotation.  Making use of the
redundant U(2) freedom noted above, we can take the U(1) rotation to be 
about the $\tau_3$ direction and fix the $\hat \tau_j^L$ and $\hat \tau_j^R$
to be
\begin{eqnarray}
    \hat \tau_j^L &=& \left\{e^{-i\psi \tau_3} \tau_1 e^{i\psi \tau_3}, \,
       e^{-i\psi \tau_3} \tau_2 e^{i\psi \tau_3}, \, \tau_3 \right\}  \cr
    \hat \tau_j^R &=& \left\{e^{i\psi \tau_3} \tau_1 e^{-i\psi \tau_3}, \,
       e^{i\psi \tau_3} \tau_2 e^{-i\psi \tau_3}, \, \tau_3 \right\} 
\end{eqnarray}
Although rotation of the relative phase $\psi$ is not an isometry,
there is a $Z_2$ symmetry that reverses its sign.  We can require
invariance under this symmetry as well, and set
$\psi=0$.\footnote{Invariance under this $Z_2$ symmetry could also be
achieved by setting $\psi=-\pi/2$; we will not explore this
possibility here.}

If we now set $T_0^L = T_0^R= 0$ and write
$T_4 = p +q\tau_3$, the Nahm matrices on the left and right intervals
then become
\begin{eqnarray}
     T_j^L(s) &=& \left[{1\over 2}\,g_1^L(s) 
         \tau_1 , \, 
    {1\over 2}\,g_1^L(s) \tau_2 , \,
                {1\over 2}\,g_3^L(s) \tau_3 + Z_L {\rm I}_2\right]   \, ,
                 \cr
     T_j^R(s) &=& \left[{1\over 2}\,g_1^R(s) 
         \tau_1 , \, 
      {1\over 2}\,g_1^R \tau_2(s), \, 
                {1\over 2}\,g_3^R(s) \tau_3 +   Z_R {\rm I}_2\right]  \, .
\label{axialT}
\end{eqnarray}
with 
\begin{eqnarray}
     g_1^L(s) &=& \cases{ -D_L \csc[D_L(s-s_L)]  \, ,&  $\kappa_L =0$   \cr
                      -D_L \cosech[D_L(s-s_L)]  \, , & $\kappa_L =1$ } \cr \cr
     g_3^L(s) &=&  \cases{ -D_L \cot[D_L(s-s_L)]  \, ,&  $\kappa_L =0$   \cr
                      -D_L \coth[D_L(s-s_L)]  \, , & $\kappa_L =1$ }  \cr\cr
     g_1^R(s) &=& \cases{ D_R \csc[D_R(s_R-s)]  \, ,&  $\kappa_R =0$   \cr
                      D_R \cosech[D_R(s_R-s)]  \, , & $\kappa_R =1$ } \cr \cr
     g_3^R(s) &=&  \cases{ D_R \cot[D_R(s_R-s)]  \, ,&  $\kappa_R =0$   \cr
                      D_R \coth[D_R(s_R-s)]  \, , & $\kappa_R =1$   . }
\label{axialfunctions}
\end{eqnarray}

We can further simplify matters by requiring that the conserved
charges from the unbroken U(1)$\times$SU(4)$\times$U(1) symmetry all
vanish.
One's first thought might be that the phases associated with these
vanishing charges could be simply dropped from the Lagrangian.  This
is not so, because there are couplings between the angular
velocities $\omega^i$ of these phases and the six non-phase
moduli ($D_L$, $D_R$, $p$, $q$, $Z_L$, and
$Z_R$) that remain after our symmetry constraints are imposed.  
If we denote the latter moduli by $y^a$, the moduli space Lagrangian can 
be written as
\begin{equation}
    L_{\rm MS} = {1\over 2} C_{ab} \,\dot y^a \dot y^b + B_{ai}\,\dot y^a \omega^i 
          + {1\over 2} E_{ij}\,\omega^i \omega^j  \, ,
\label{LMS}
\end{equation}
where the metric coefficients $C_{ab}$, $B_{ai}$, and $E_{ij}$  depend only on
the $y^a$.  By means of a Legendre transformation we can convert this to an
effective Lagrangian in which the dependence on the $\omega^i$ is replaced by 
a dependence on the conserved charges
\begin{equation}
    Q_j = E_{ij}\omega^i + B_{ja}\dot y^a  \, .
\end{equation}
If all of the $Q_j$ vanish, this effective Lagrangian reduces to 
\begin{equation}
     L_{\rm MS,\,eff} = {1\over 2}\left[ C_{ab} - B_{ai} E^{-1}_{ij} B_{jb} \right]
 \dot y^a \dot y^b  \, .
\label{LMSeffDef}
\end{equation}

As we did for the (1,[1],1) example, we will take advantage of the
isometries of the moduli space and calculate the metric at the point
$V=I$.  We start our calculation by displaying the Nahm data.
The $T^L_\mu(s)$ and $T^R_\mu(s)$, as well as $T_4$, were given 
above.
With $V=I$, $A=K^{1/2}$, where
\begin{equation}
      K = \left(\matrix{p+q+C+R  & 0 & 0 & 0 \cr \cr
                  0 & p-q-C+R & 2B  & 0 \cr \cr
                  0 & 2B & p+q-C-R & 0 \cr \cr 
                 0 & 0 & 0 & p-q+C-R } \right)
\label{Kdisplay}
\end{equation}
with 
\begin{eqnarray}
      B &\equiv& {1\over 2}\left[ g_1^L(s_0) - g_1^R(s_0)\right] \, , \cr\cr
      C &\equiv& {1\over 2}\left[g_3^L(s_0) - g_3^R(s_0)\right] \, ,  \cr\cr
      R &\equiv& Z_L - Z_R     \, .
\end{eqnarray}
\pagebreak
Note that $K$ has been written so that the Greek indices in
Eq.~(\ref{KmatrixDef}) label $2\times 2$ blocks; the elements within
each block are labeled by the indices $r$ and $s$.

Given this Nahm data, the calculation of $L_{\rm MS,\,eff}$ can be
organized as follows:

1) {\bf Calculate the derivatives of the Nahm data with respect to the
$\bm{y^a}$.}  On the left and right intervals the only nonvanishing
derivatives are those of the $T_\mu$ with respect to the corresponding
$D$ and $Z$, but $A$ has nonzero derivatives with respect to all of the 
$y^a$.  The calculation of these is somewhat involved, and so we
relegate it to the Appendix.  The explicit form of the results are actually 
not needed until the final step 6.

2) {\bf Determine whether the tangent vectors obtained in step 1 require
any additional gauge actions to put them into background gauge.}  It 
is easy to see that the derivatives of the $T_\mu$ on the left and right
intervals obey Eq.~(\ref{backgroundgauge}).  
The pieces arising from the data at
$s_0$ require a bit more care.  With $V$ taken to be the identity, $A$
is Hermitian.  Equation~(\ref{jumpgaugecondition}) then implies that a
compensating gauge action is only needed if
\begin{equation}
    \left[ {\partial A \over \partial y^a} , A \right]_{\alpha r,\alpha s}
          \ne 0
\end{equation}
for any coordinate $y^a$.  To see that this quantity always vanishes,
first note that both $A$ and $\partial A/\partial y^a$ have the same
block diagonal form as $K$, and that a nonvanishing commutator can
only arise from the middle $2\times 2$ block.  Within this block, the
matrices are all linear combinations of the identity and the Pauli
matrices $\rho_x$ and $\rho_z$.  Any commutator must then be
proportional to $\rho_y$, and thus would not contribute after the
trace over $\alpha$ was taken.\footnote{This cancellation is a
consequence of the axial symmetry, because otherwise there is also a
$\rho_y$ contribution to $K$.}   

3) {\bf Determine which $\bm{B_{ja}}$ are nonvanishing.}  Because the
tangent vectors for the $y^a$ do not require compensating gauge
actions, $B_{ja}$ is given just by the first term on the last line of
Eq.~(\ref{jumpmetric}).  This gives
\begin{equation}
    B_{ja} =  -{2\pi i}\, \Tr \left( 
        \left[ {\partial A \over \partial y^a} , A \right] t_j \right)  \, ,
\end{equation}
where $t_j$ is the Hermitian generator corresponding to the $j$th phase.
From the remarks of the previous paragraph, we see that we can choose
the $t_j$ so that the only nonzero $B_{ja}$ come from the generator
that has a $\rho_y$ in the middle $2\times 2$ block and zeros elsewhere; we label
this generator $t_2$.

4) {\bf Show that the tangent vector corresponding to the U(4) action
generated by $\bm{t_2}$ does not need a compensating gauge action.}
Referring to Eq.~(\ref{jumpgaugecondition}), we see that this is
equivalent to showing that
\begin{equation}
   0 = \left(A t_2 A \right)_{\alpha r, \alpha s}
\end{equation}
for all values of $r$ and $s$.  It is easy to verify that this follows
from the symmetric block diagonal form of $A$.

5) {\bf Calculate $\bm{E^{-1}_{22}}$. }  Because the $B_{ja}$ vanish if $j\ne 2$,
we only need this one element of the matrix $E^{-1}$.  Using the fact that 
the $t_2$ tangent vector 
has no compensating gauge action, Eq.~(\ref{jumpmetric}) gives
\begin{equation}
   E_{2j} =  2\pi \, \Tr ( A\{t_2,t_j \} A )
          = 2\pi \, \Tr (K\{t_2,t_j \})   \, .
\end{equation}
This vanishes unless $j=2$, implying that 
\begin{equation} 
    E^{-1}_{22} = \left( E_{22} \right)^{-1} = {1 \over 8\pi (p-C) }  \, .
\label{Etwotwo}
\end{equation}

6) {\bf Evaluate the $\bm{C_{ab}}$ and the $\bm{B_{2a}}$ and substitute the
  results into Eq.~(\ref{LMSeffDef}) to obtain $\bm{ L_{\rm MS, eff}}$.}
The details of this are given in the Appendix.  Instead of writing the 
result directly in terms of the $y^a$, it is more convenient to express 
it in terms of the four eigenvalues of $K$,
\begin{eqnarray}
    \lambda_1 &=& p + q + R + C  \, ,  \cr
    \lambda_2 &=& p - q - R + C  \, ,  \cr
    \lambda_+ &=& p - C + \sqrt{4B^2 +(q-R)^2}  \, ,    \cr
    \lambda_- &=& p - C -\sqrt{4B^2 +(q-R)^2}  \, ,
\end{eqnarray}
and the variable
\begin{equation}
     \theta = \tan^{-1}\left({2B \over R-q} \right)  \, .
\end{equation}
We can then write
\begin{equation} 
   L_{\rm MS, eff}  =  M_L\, \dot Z_L^2 + M_R \, \dot Z_R^2  
     + {1 \over 2}I_{DD}^L \, \dot D_L^2 + {1 \over 2} I_{DD}^R \, \dot D_R^2  
      + {\pi \over 2} \sum_\sigma  { \dot\lambda_\sigma^2 \over \lambda_\sigma} 
     + {\pi \over 2}
     {\left(\lambda_+  - \lambda_- \right)^2
           \over \left(\lambda_+ + \lambda_- \right)} \dot\theta^2   \, ,
\label{twotwotwoLag}
\end{equation}
where $I_{DD}$ is the function given in Eq.~(\ref{IDDdef}).  (Of course, when
obtaining the equations of motion from this Lagrangian one must remember that the
$\lambda_\sigma$ and $\theta$ are not independent variables.)

\subsection{The large-mass limit}

Considerable simplification can be achieved by working in the
``large-mass limit'' in which the massive monopole core radii,
$M_L^{-1}$ and $M_R^{-1}$, are much less than all other relevant
distance scales.\footnote{It must be kept in mind that this limit
involves a comparison between the monopole masses and the cloud sizes
and massive monopole separations.  While the masses are, of course,
constant, the evolution of the other quantities may invalidate this
limit at large times.  This would happen, for example, in a geodesic
motion that started with a large-mass $\kappa=0$ Dancer solution,
passed through the spherically symmetric point where the symmetry axes
in Fig.~\ref{dancerspace} meet, and then moved out toward the
large-mass $\kappa=1$ solutions.}  There are four possible cases,
depending on the values of $\kappa_L$ and $\kappa_R$.  We will examine
the two with $\kappa_L = \kappa_R$.

\subsubsection{Hyperbolic solutions, $\kappa_L = \kappa_R = 1$ }

Here we
take $\mu_L= M_LD_L/4\pi$ and $\mu_R = M_R D_R/4\pi$ both large, with $D_L$ and
$D_R$ held fixed.  In this limit the Dancer clouds have minimum size and
$D_L$ and $D_R$ are the separations between the massive monopoles of
the same species.  Up to exponentially small corrections,
\begin{equation}
    B=0\, , \qquad       C = -{1\over 2}(D_L + D_R)  \, .
\end{equation}   
Substituting these values, as well as the asymptotic values of
$I_{DD}^L$ and $I_{DD}^R$ from Eq.~(\ref{IDDhyperlimit}), into
Eq.~(\ref{twotwotwoLag}) gives
\begin{eqnarray}
      L_{\rm MS, eff}  &=& {1 \over 2} M_L \, \dot Z_1^2 
         + {1 \over 2} M_R \,\dot Z_4^2
    +   {2\pi (Z_1 -Z_4) \over [( p- q)^2 - ( Z_1 - Z_4)^2]}
            \, (\dot p-\dot q) \, (\dot Z_1 -\dot Z_4)
    \cr\cr &+&
    {\pi (p-q) \over [(p-q)^2 - (Z_1 -Z_4)^2]}
              \left[(\dot p-\dot q)^2 + (\dot Z_1 -\dot Z_4)^2 \right]
  \cr\cr &+&
         {1 \over 2} M_L \, \dot Z_2^2
         + {1 \over 2} M_R \, \dot Z_3^2
        +   {2\pi(Z_2 -Z_3) \over [(p+q)^2 - (Z_2 -Z_3)^2]}
            \, (\dot p+\dot q) \, (\dot Z_2 -\dot Z_3)
          \cr\cr &+&
        {\pi(p+q) \over [(p+q)^2 - (Z_2 -Z_3)^2]}
              \left[(\dot p+\dot q)^2 + (\dot Z_2 -\dot Z_3)^2 \right]  \, ,
\label{hypermetric}
\end{eqnarray}
where
\begin{eqnarray}
      Z_1 &=& Z_L + {D_L\over 2} \, ,  \cr \cr
      Z_2 &=& Z_L - {D_L\over 2} \, ,  \cr \cr
      Z_3 &=& Z_R + {D_R\over 2}  \, , \cr \cr
      Z_4 &=& Z_R - {D_R\over 2}     \, .
\end{eqnarray}

Examination of Eq.~(\ref{hypermetric}) shows that the metric is the
sum of two independent pieces, one involving $Z_1$, $Z_4$, and $p-q$,
and one involving $Z_2$, $Z_3$, and $p+q$.  Each of these describes a
(1,[1],1) SU(4) system.  [Indeed, this could have been foreseen by
recalling the results of Ref.~\cite{Houghton:2002bz}, where it was
shown that the SU(6) solutions with two minimal Dancer clouds and all
SU(2) orientations aligned were essentially superpositions of two
independent SU(4) (1,[1],1) solutions.]  The splitting of the metric
here implies that the two nontrivial clouds are completely decoupled
from each other.  Hence, this limiting case does not shed light on the
interactions between clouds, which is our primary interest in this
paper.  We therefore turn to the second limiting case.

\subsubsection{Trigonometric solutions, $\kappa_L = \kappa_R=0$}

For these, we
take $\mu_L = (s_0-s_L)D_L= M_LD_L/4\pi$ and 
$\mu_R =(s_R-s_0)D_R = M_R D_R/4\pi$ to be just less than the
maximum allowed value, $\pi$.  In this regime the approximate radius
of the Dancer cloud is
\begin{equation}
     a = {D \over 2( \pi - \mu)} \gg M^{-1}  \, .
\end{equation}
To leading order, then, we can write
\begin{equation}
     C = -B = (a_L + a_R) \equiv\tilde a  \, .
\end{equation}
In addition, using Eq.~(\ref{IDDtriglimit}), we find, again to leading
order, that 
\begin{equation}
     I_{DD}^L \, dD_L^2 + I_{DD}^R \, dD_R^2  = 
      4\pi \,{da_L^2 \over  a_L} + 4\pi \, {da_R^2 \over  a_R}   
       = 16\pi  \left(d\sqrt{\tilde a}\right)^2 + 16 \pi \,\tilde a d \phi^2  \, ,
\end{equation}
where $\phi = \tan^{-1}(\sqrt{a_L/a_R})$.  

We can take the center of mass, $M_LZ_L + M_RZ_R$, to be at 
rest and define a reduced 
mass ${\cal M} = M_L M_R/(M_L + M_R)$.  The effective moduli space 
Lagrangian of Eq.~(\ref{LMSeffDef}) then reduces to
\begin{equation}
    L_{\rm MS, eff}  =  {\cal M}\, \dot R^2
     +  {\pi\over 2} \sum_\sigma  { \dot\lambda_\sigma^2 \over \lambda_\sigma}
      + {\pi \over 2}
     {\left(\lambda_+  - \lambda_- \right)^2
           \over \left(\lambda_+ + \lambda_- \right)} \dot\theta^2
   + 4\pi \, {\dot{\tilde a}^2 \over \tilde a} + 16\pi\, \tilde a \, \dot\phi^2 \, .
\label{effLag}
\end{equation}
Note that, except in the $\dot \phi^2$ term, the Dancer cloud size parameters
$a_L$ and $a_R$ only enter the Lagrangian through their sum $\tilde a$.  This is
a consequence of our having aligned the U(1) phases of the two Dancer
clouds, as described in Sec.~\ref{cylindrical}.

Finally, in the limit of large monopole mass we can treat $R = Z_L
-Z_R$ as being constant in time, and so drop the first term on the right-hand side
of Eq.~(\ref{effLag}).  For the sake of simplicity, we will set $R=0$.  In the 
large-mass limit in which we are working, this makes the system essentially spherically
symmetric.   It also sets $\theta = - \tan^{-1}(2\tilde a/q)$.

\section{Cloud dynamics}
\label{dynamics}

We now focus on the dynamics of the trigonometric solutions discussed at the end
of the previous section.  We work in the large-mass limit with $R=0$.
The eigenvalues $\lambda_\sigma$ of the matrix $K$ are then 
\begin{eqnarray}
    \lambda_1 &=& p + q +\tilde a  \, ,  \cr
    \lambda_2 &=& p - q +\tilde a  \, ,  \cr
    \lambda_+ &=& p -\tilde a + \sqrt{q^2 + 4 \tilde a^2}  \, ,    \cr
    \lambda_- &=& p - \tilde a -\sqrt{q^2 + 4 \tilde a^2}  \, ,
\end{eqnarray}
As noted in Sec.~\ref{Nahmsec}, these eigenvalues must all be
positive.  Applying this constraint to the smallest eigenvalue,
$\lambda_-$, gives the inequality\footnote{For a fixed static
solution, $q$ is naturally defined to be positive. However,
when describing time-dependent solutions it is convenient to allow
$q$ to change sign when it goes through a zero.}
\begin{equation}
       p - |q| \ge \tilde a \ge 0   \, .
\label{lambdabound}
\end{equation}

\subsection{Asymptotic behavior}

The system is particularly easy to analyze at large times (either positive 
or negative).   The eigenvalues are then all large, with
$p\pm q \gg \tilde a= a_L + a_R$, and 
\begin{eqnarray} 
      \lambda_1 &\approx& \lambda_+ \approx p + q  \, , \cr
      \lambda_2 &\approx& \lambda_- \approx p - q   \, .
\end{eqnarray}
Substituting these into Eq.~(\ref{effLag}), and noting that the $\dot
\theta^2$ term in the Lagrangian is suppressed, we see that the
dynamics is well described by the Lagrangian
\begin{equation}
    L_{\rm asym} = \pi {(\dot p + \dot q)^2\over p+q}
       +  \pi {(\dot p - \dot q)^2\over p-q}
       + 4\pi {\dot a_L^2 \over a_L} 
       +  4\pi {\dot a_R^2 \over a_R} \, .
\end{equation}

This can be viewed as describing a system composed of four
noninteracting spherical clouds: two ``SU(4) clouds'', with cloud
parameters $(p+q)$ and $(p-q)$, and two Dancer clouds, with cloud
parameters $a_L$ and $a_R$.  (We will refer to these cloud parameters
as radii, but it should be kept in mind that the cloud structure does
not allow a precise and unambiguous definition of its radius.)
These evolve according to
\begin{eqnarray}
     p\pm q &=&  {1\over 2} C_\pm(t - t_\pm)^2  \, ,  \cr
     a_{L,R} &=& {1\over 2} C_{L,R}(t - t_{L,R})^2   \, , 
\end{eqnarray}
where the $C_i$ and $t_i$ are arbitrary constants.\footnote{These
formulas imply that at very large times the clouds would be expanding
at speeds greater than that of light.  A more detailed analysis of
cloud behavior~\cite{Chen:2001qt} shows that at these times the moduli
space approximation breaks down, and that instead the cloud expansion
is best described as a wavefront moving at the speed of light.}  The
total energy is divided into four separately conserved parts,
\begin{eqnarray}
   E_{p+q} &=&  \pi {(\dot p + \dot q)^2\over p+q} = \pi C_+  \, , \cr 
   E_{p-q} &=& \pi {(\dot p - \dot q)^2\over p-q} =\pi C_-  \, , \cr
   E_L &=& 4\pi {\dot a_L^2 \over a_L} = 4\pi C_L  \, , \cr
   E_R &=& 4\pi {\dot a_R^2 \over a_R} = 4\pi C_R \, .
\end{eqnarray}

Note that this asymptotic separation into noninteracting clouds did
not require that $(p+q)$-cloud and the $(p-q)$-cloud be very
different in size, but only that they both be much larger than the
Dancer clouds.  This can be understood by recalling the description of
the corresponding static solutions in Ref.~\cite{Houghton:2002bz}.  By
analyzing the magnetic field in the regions between the cloud radii,
it was found that the non-Abelian part of the effective magnetic
charge, $Q_{\rm NA}$, in each of the regions can be diagonalized,
with\footnote{We have arbitrarily chosen $q>0$ and $a_R > a_L$.}
\begin{equation}
      Q_{\rm NA} = \cases{ {\rm diag(0,0,0,0)} \, , & $r \gg p+q$ \cr
             {\rm diag(0,-1,0,1)} \, , & $p+q \gg r \gg p-q$ \cr
             {\rm diag(-1,-1,1,1)} \, , & $p-q \gg r \gg a_R$ \cr
             {\rm diag(-2,0,1,1)} \, , & $a_R \gg r \gg a_L$ \cr       
             {\rm diag(-2,0,0,2)} \, , & $a_L \gg r $ . }   
\end{equation}
In other words, the clouds act as if they have magnetic charges
\begin{eqnarray}
     Q_{p+q} &=& {\rm diag(0,1,0,-1)}    \, , \cr
     Q_{p-q} &=& {\rm diag(1,0,-1,0)}    \, , \cr
     Q_{a_R} &=& {\rm diag(1,-1,0,0)}    \, , \cr
     Q_{a_L} &=& {\rm diag(0,0,1,-1)}   \, . 
\end{eqnarray}
Thus, the $(p+q)$- and the $(p-q)$-clouds lie in mutually commuting
SU(2) subgroups of the unbroken SU(4), and so can only affect each
other via interactions mediated by one or both of the Dancer clouds.
When $p \pm q \gg \tilde a$, these interactions are negligible, in
accordance with the discussion in Sec.~\ref{largeCloudSec}.
Similarly, the two Dancer clouds decouple from each other in this
asymptotic regime, regardless of their relative sizes.

\subsection{Scattering}

We have a system of four clouds that are asymptotically
noninteracting.  The asymptotic solutions indicate that they are all
contracting at large negative times, and expanding at large positive
times.  Their interactions at intermediate times can be viewed as a
series of one or more scattering processes.  These can be studied by
starting with an initial configuration containing well-separated
clouds and then, using numerical simulations, letting the system
evolve under the equations of motion that follow from the Lagrangian
of Eq.~(\ref{effLag}).

\begin{figure}
\centering
\begin{tabular}{cc}
\epsfig{file=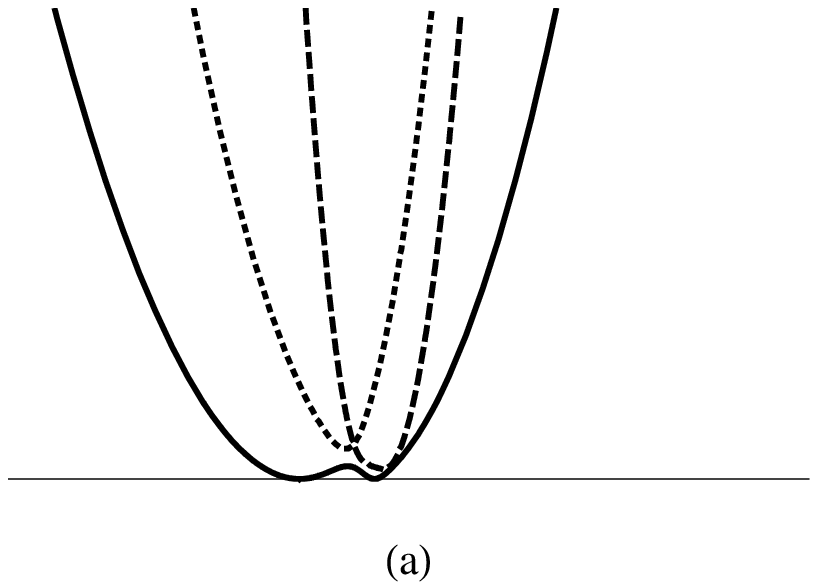,width=0.49\linewidth,clip=} & 
\epsfig{file=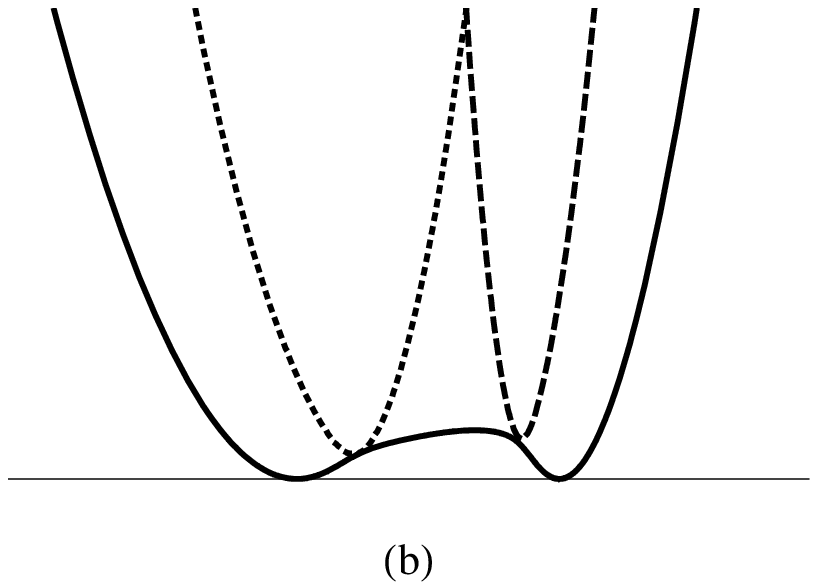,width=0.49\linewidth,clip=} \\
\end{tabular}
\caption{Two examples of cloud collisions.  The horizontal axis
represents time, and the vertical axis cloud size.  The incoming $p-q$
and $p+q$ clouds, represented by the dotted and dashed lines
respectively, collide with the Dancer cloud (solid line) and then
expand to infinity.  }
\label{multicloudscatter}
\end{figure}

We show two typical examples of this in Fig.~\ref{multicloudscatter}.
Both of these simulations were performed with the constant of motion
$J={\tilde a}^2\dot\phi$ set equal to zero, so that the ratio of the
Dancer cloud radii remains constant throughout.  The evolution does
not depend on this ratio, but only on the sum of the Dancer radii,
$\tilde a$, which is shown by the solid line in these plots.  There is
some ambiguity in defining the size of the two SU(4) clouds [e.g., the
differences between $\lambda_1$, $\lambda_+$, and $p+q$ are negligible
at large times, but not necessarily when the SU(4) and Dancer clouds
are comparable in size].  We have, somewhat arbitrarily, chosen to
plot $p+q$ (dotted line) and $p-q$ (dashed line).

These plots show several features, common to all of the examples that
we have examined, that should be noted.  First, the SU(4) clouds
always remain larger than the Dancer clouds (in fact, larger than the
sum of the Dancer radii), as should be expected from the bound in
Eq.~(\ref{lambdabound}).  In the asymptotic solutions, the SU(4) cloud
radii have parabolic dependences on time, with a minimum radius of
zero.  In the actual interacting solutions, their behavior is rather
similar, except that the vertex of the parabola is raised so that it
occurs at or near the point when the SU(4) cloud radius is equal to
$\tilde a$.  (Given the ambiguity in defining the cloud radii, the
distinction between exact coincidence of these values, as in
Fig.~\ref{multicloudscatter}a, or a slight gap between them, as in
Fig.~\ref{multicloudscatter}b, is not meaningful.)  In particular, the
overlap (or near overlap) between the SU(4) clouds and the Dancer
clouds is relatively brief, suggestive of a rather short and sharp
interaction.

Also, from examining simulations for a variety of initial conditions,
we see that, just as in the asymptotic limit, the $(p+q)$- and
$(p-q)$-clouds do not appear to interact directly with each other.
This suggests that we focus on the interaction of just one of these
SU(4) clouds with the Dancer clouds.  We can do this by choosing
initial conditions such that the $(p+q)$-cloud is very large (and
therefore essentially not interacting with the rest of the system) at
the time that the $(p-q)$- and Dancer clouds interact.  In fact, we
can simplify our analysis by taking the $(p+q)$-cloud to be at
infinity; i.e., by taking the limit $p \to \infty$, with $\dot p^2/p$
and $\delta \equiv p-q$ held fixed.  In this limit $\lambda_1 =
2p-\delta +\tilde a$ and $\lambda_+ = 2p-\delta - \tilde a +O(1/p)$
tend to infinity, while
\begin{eqnarray}
     \lambda_2 &=& \delta + \tilde a  \, ,  \cr
     \lambda_- &=& \delta - \tilde a  \, .
\end{eqnarray}
If we drop the terms proportional to $\dot p^2$ that decouple from
everything else, and restrict ourselves to the $J=0$ case where the
ratio of $a_L/a_R$ remains constant, the effective Lagrangian of
Eq.~(\ref{effLag}) reduces to
\begin{equation}  
     L_{\rm MS, red} 
    = {\pi \over 2} {(\dot \delta + \dot{\tilde a})^2 \over (\delta +\tilde a)}
    + {\pi \over 2} {(\dot \delta - \dot {\tilde a})^2 \over (\delta -\tilde a)}
    + 4\pi {\dot {\tilde a}^2 \over \tilde a} \, .
\label{aDeltaLag}
\end{equation}

Keeping in mind that we expect $\delta \gg \tilde a$ at large times, and noting that
this purely kinetic Lagrangian is equal to the energy, we can write 
\begin{eqnarray}
   E  &=& \left[{\pi\delta {\dot \delta}^2 \over {\delta^2-\tilde a}^2}
        - {\pi \tilde a \dot{\tilde a}\dot\delta\over {\delta^2-\tilde a}^2}\right] 
      + \left[{4\pi{\dot {\tilde a}}^2 \over \tilde a} 
       + {\pi \delta {\dot {\tilde a}}^2 \over {\delta^2-\tilde a}^2}
    - {\pi \tilde a \dot{\tilde a}\dot\delta\over {\delta^2-\tilde a}^2}\right]
   \cr
    &\equiv& E_\delta + E_a  \, .
\end{eqnarray}
where we have defined SU(4) and Dancer cloud energies whose asymptotic values
at large $|t|$ are
\begin{eqnarray}
    E_\delta &=& \pi \,  {\dot \delta^2 \over \delta}  \, ,  \cr
    E_a  &=& 4 \pi \, {\dot {\tilde a}^2 \over \tilde a}  \, .
\label{asymE} 
\end{eqnarray}
It follows that the trajectories at large
negative times, when $\delta \gg \tilde a$, are of the form
\begin{eqnarray}
   \delta (t) &=&  { E_\delta \over 4\pi} \, (t-t_\delta)^2 \, , \cr\cr 
   \tilde a(t) &=& { E_a \over 16\pi} \, (t-t_a)^2     \, .
\label{asymInitCond}
\end{eqnarray}
The trajectories at large positive times are of the same form,
except that the values of the various constants of motion are
changed as a result of the interactions between the clouds.

\begin{figure}
\centering
\begin{tabular}{cc}
\epsfig{file=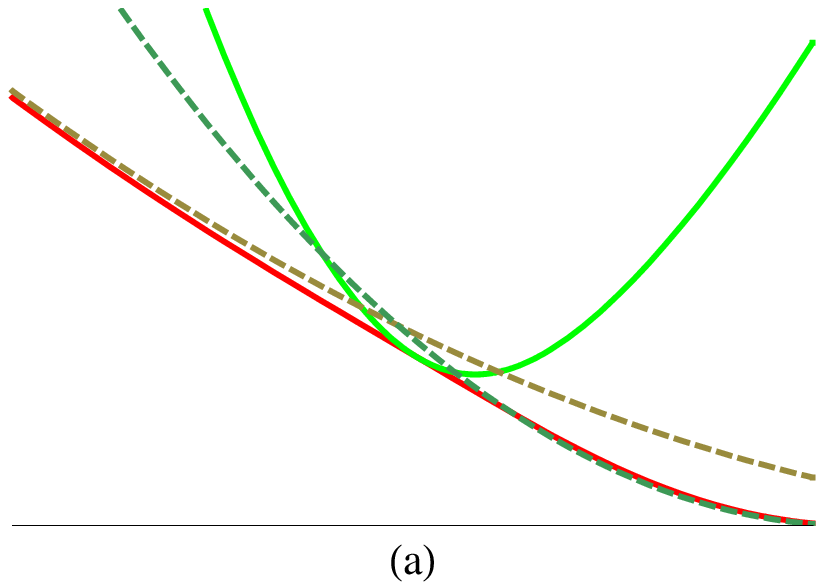,width=0.49\linewidth,clip=} & 
\epsfig{file=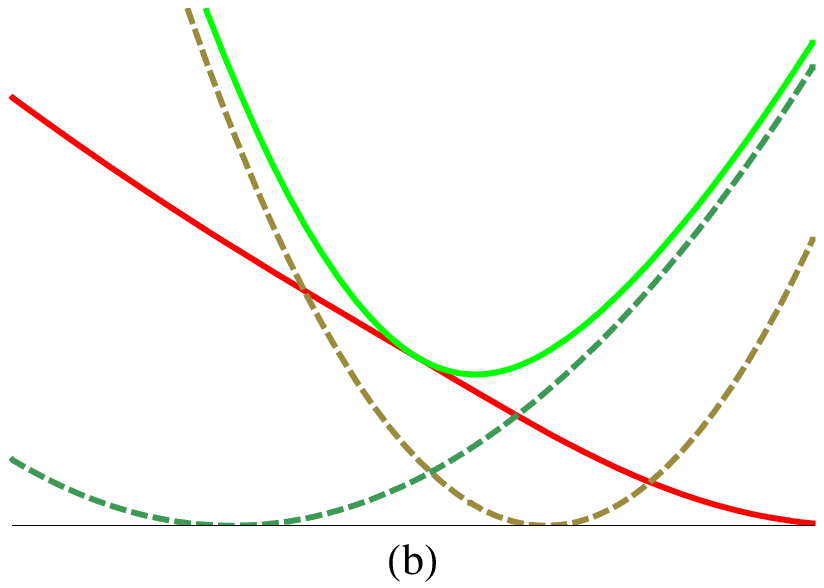,width=0.49\linewidth,clip=}  
\end{tabular}
\caption{A typical collision between the Dancer cloud and an SU(4)
cloud.  The horizontal axis
represents time, and the vertical axis cloud size.
The actual cloud trajectories are shown as solid lines.  In
(a) the dashed lines indicate the initial and final asymptotic
trajectories of the Dancer cloud, while in (b) they indicate the 
asymptotic trajectories of the SU(4) cloud.}
\label{asymcurves}
\end{figure}

The form of Eq.~(\ref{aDeltaLag}) is strikingly similar to that of
Eq.~(\ref{su4Lag}), with the Dancer cloud parameter $\tilde a$ playing
a similar role to $R$, the separation between the massive monopoles in
the SU(4) (1,[1],1) solution, and the fixed monopole reduced mass
$\mu$ replaced by the variable $\tilde a^{-1}$.  This seems
surprising, since the previous case involved massive monopoles hitting
an ellipsoidal cloud at two distinct points, while in the present case
two nested spherical clouds are meeting each other at all points.
Nevertheless, the similarity in the Lagrangians suggests that the
interactions should be similar.  In particular, the analysis of the
SU(4) dynamics in Ref.~\cite{Chen:2001ge} found that the interaction
between the cloud and the massive monopoles was relatively brief,
taking place over a distance of order $\mu^{-1}$.  This suggests
similarly brief interactions in the present case, with the interaction
largely restricted to the time when $\delta - \tilde a$ is itself of
order $\tilde a$.  We saw some indication of this, with all of the
clouds present, in Fig.~\ref{multicloudscatter}.  We illustrate this
more clearly in the two-cloud case in Fig.~\ref{asymcurves}, where we show the
transition from the initial asymptotic trajectories to the final ones.

Equation~(\ref{asymInitCond}) suggests that an arbitrary solution
depends on four initial constants, $t_a$, $t_\delta$, $E_a$, and
$E_\delta$.  It is clear that time-translation invariance can be used
to eliminate one of these.  In addition, the Lagrangian of
Eq.~(\ref{aDeltaLag}) has some interesting scaling properties.  The
only effect of the rescalings
\begin{eqnarray}
      \tilde a &\rightarrow& \tilde a' = \lambda \tilde a \, , \cr
      \delta &\rightarrow& \delta' = \lambda \delta \, ,\cr
       t &\rightarrow& t' = \kappa t  
\end{eqnarray}
is to multiply the Lagrangian by an overall factor of
$\lambda/\kappa^2$.  Hence, given any solution of the equations of
motion, these rescalings will generate a two-parameter set of
solutions.  Thus, to study the full range of possible solutions we
really only need to vary a single continuous parameter, which we
choose to be $E_a/E_\delta$.  (Note that applying the constraint
$\delta > \tilde a$ in the asymptotic region implies that
$E_a/E_\delta <4$.)  Also, since the rescaling cannot reverse the
time ordering, we must consider separately the cases $t_a -t_\delta
>0$ and $t_a - t_\delta <0$.

\begin{figure}[t]
\centering
\begin{tabular}{cc}
\epsfig{file=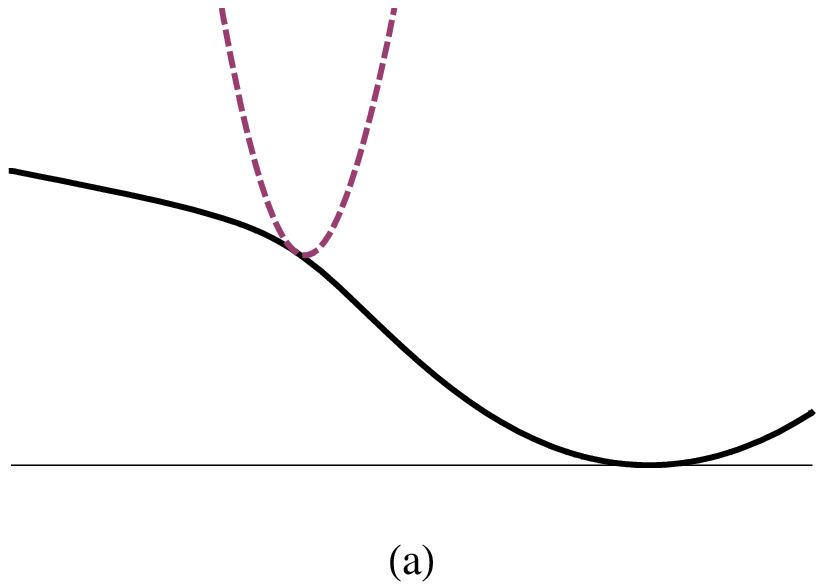,width=0.49\linewidth,clip=} & 
\epsfig{file=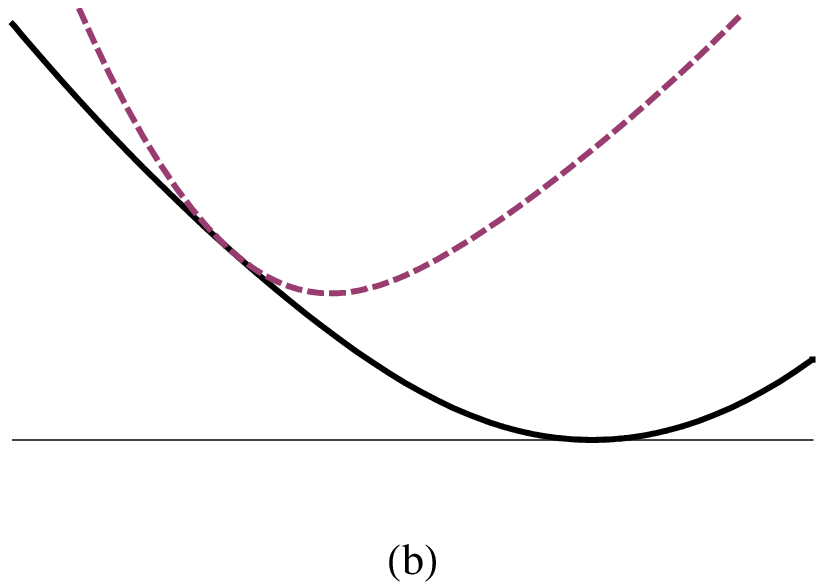,width=0.49\linewidth,clip=} \\
\epsfig{file=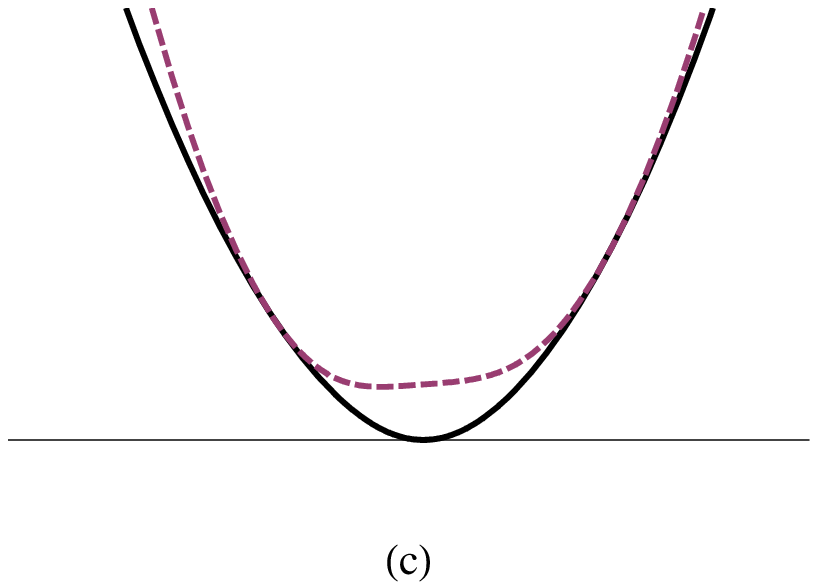,width=0.49\linewidth,clip=} &
\epsfig{file=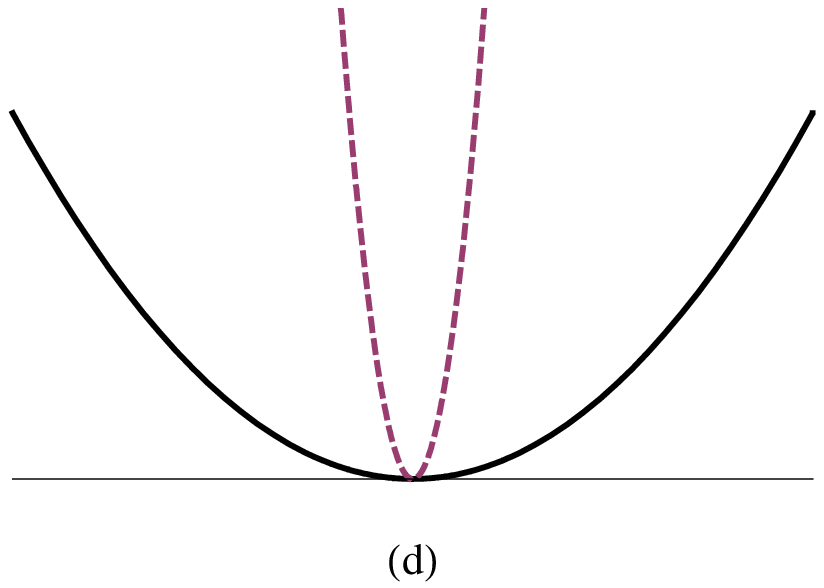,width=0.49\linewidth,clip=} \\
\epsfig{file=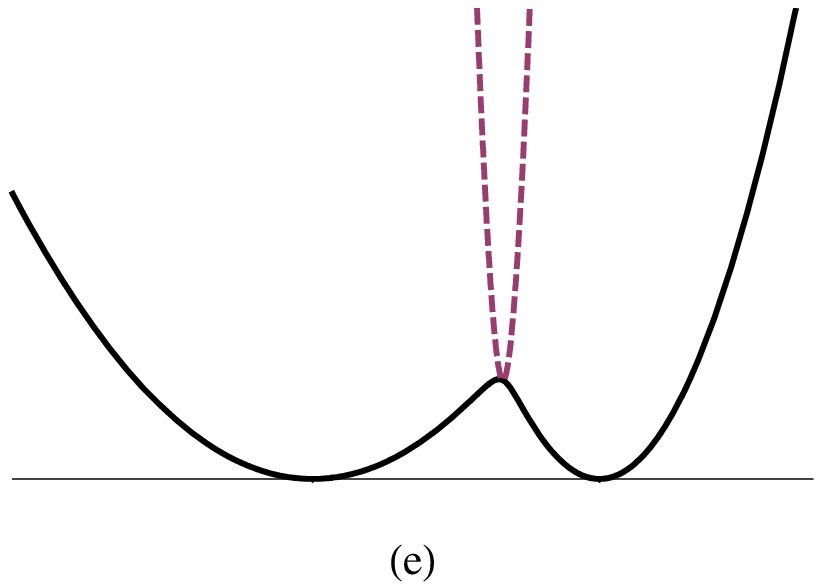,width=0.49\linewidth,clip=} &
\epsfig{file=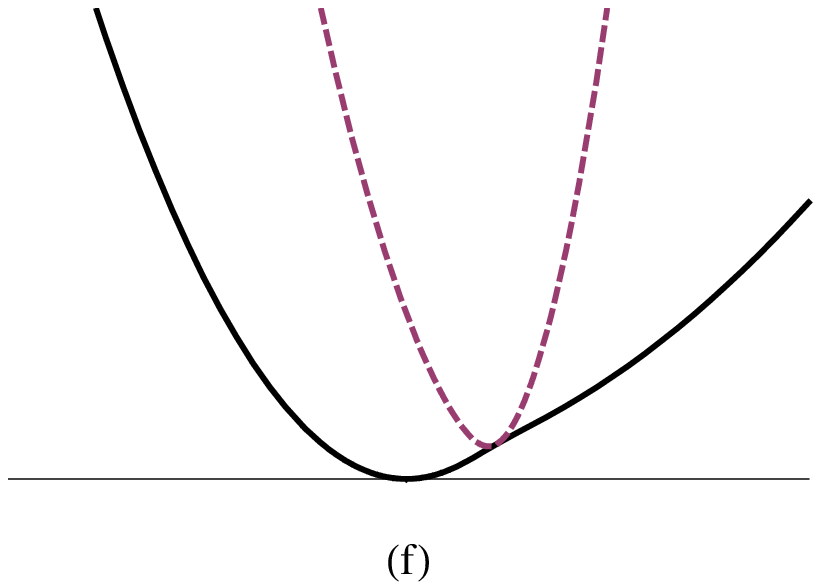,width=0.49\linewidth,clip=}
\end{tabular}
\caption{Typical interactions between a Dancer cloud (solid black line) and 
an SU(4) cloud (dashed purple line). The horizontal axis
represents time, and the vertical axis cloud size.}
\label{manyInt}
\end{figure}

The range of possibilities is illustrated in Fig.~\ref{manyInt}.  If
$t_a -t_\delta >0$, the collapsing SU(4) cloud collides with the
Dancer cloud while the latter is also collapsing.  Three examples of this 
are shown in Fig.~\ref{manyInt}a-c, with the value of $E_a/E_\delta$ increasing
from one to the next.  In all three cases the SU(4) cloud loses energy to the Dancer
cloud.  In the last case, where $E_a/E_\delta$ is initially close to its maximum
allowed value, the SU(4) cloud loses so much energy that the inequality 
$E_a/E_\delta <4$ is temporarily violated.  Because the cloud radii
both increase like $Et^2$, there must be a second interaction in which
the Dancer cloud overtakes the SU(4) cloud and transfers back enough energy
that the inequality is satisfied at large times.   The crossover from the behavior 
shown in Fig.~\ref{manyInt}b to that in Fig.~\ref{manyInt}c occurs when 
$E_a/E_\delta \approx 2$.

In the borderline case, $t_a - t_\delta =0$, the two clouds arrive at
the origin simultaneously, as shown in Fig.~\ref{manyInt}d.  In this
case the asymptotic solution of Eq.~(\ref{asymInitCond}) is exact for
all times, and no energy is exchanged between the clouds.

Finally, we come to the case where $t_a -t_\delta <0$.  Here, the
collapsing SU(4) cloud only reaches the Dancer cloud after the latter
has already reached its minimum size and begun to expand.  If
$E_a/E_\delta <4$ is sufficiently large, as in Fig.~\ref{manyInt}e,
the Dancer cloud loses some energy to the SU(4) cloud, but
continues to expand, although at a reduced speed.  (This is then a time-reversed
version of a solution with $t_a -t_\delta >0$.)  However, if
$E_a/E_\delta < 4$ is small enough, as in Fig.~\ref{manyInt}f, the
collision can reverse the expansion of the Dancer cloud and have it
shrink to zero radius a second time.  The boundary between these two regimes 
is at $E_a/E_\delta \approx 2.6$.

\begin{figure}
\centering
\begin{tabular}{cc}
\epsfig{file= 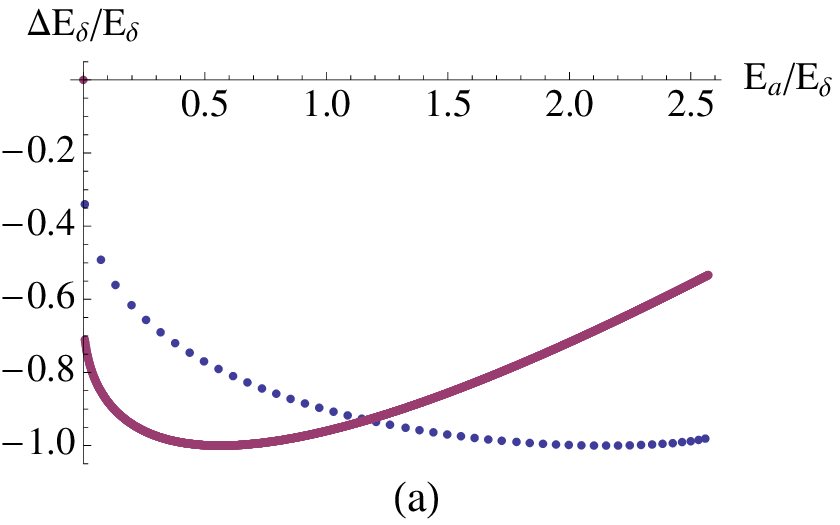,width=0.49\linewidth,clip=} &
\epsfig{file= 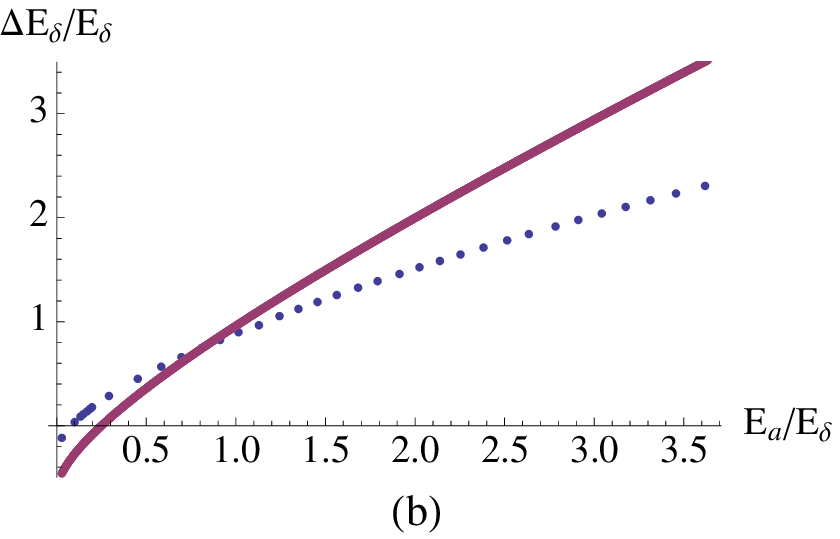,width=0.49\linewidth,clip=} \\
\epsfig{file= 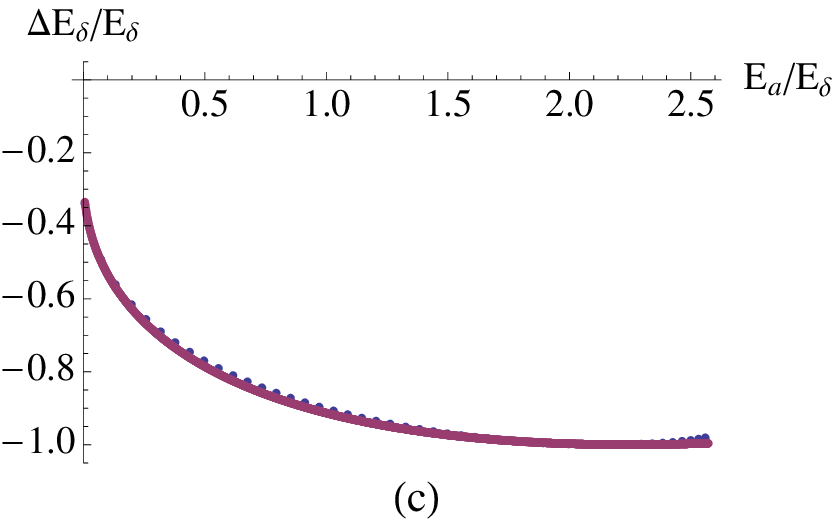,width=0.49\linewidth,clip=} &
\epsfig{file= 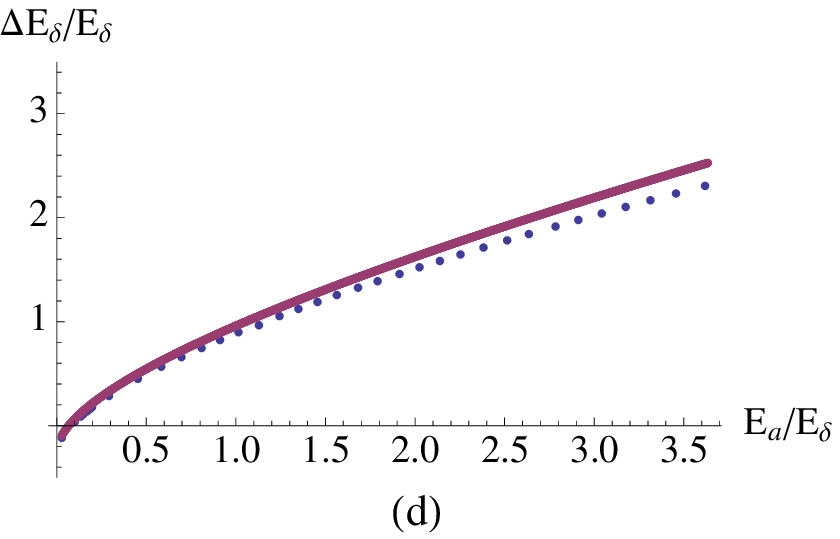,width=0.49\linewidth,clip=}
\end{tabular}
\caption{Energy change between the SU(4) and Dancer clouds.  In (a)
and (c), both clouds are contracting at the time of collision, while
in (b) and (d) a contracting SU(4) cloud collides with an expanding
Dancer cloud.  In all four plots the dotted blue curve shows the
transfer observed in the simulations.  In (a) and (b) the solid purple
curve shows the prediction for an elastic collision, while in (c) and
(d) it indicates the prediction for an inelastic collision with
$\Delta = -{1\over 5} (3 v_{\delta i}^2 + 4v_{\delta i}v_{ai} -4
v_{ai}^2)$.}
\label{energyTransFig}
\end{figure}

While these plots are sufficient to provide a qualitative
understanding of the interactions, it would be nice to have some more
quantitative results as well.  Let us first consider the energy
transferred during the collision.  The fact that the interaction
between the clouds takes place over a relatively short time interval
suggests a naive model that treats the interaction as an instantaneous elastic
collision of two rigid shells, with kinetic energy and radial momentum
($\sum_a M_a \dot r_a$) conserved.  Because the Dancer cloud
has four times the kinetic energy of the SU(4) cloud for the same value of
the velocity [see Eq.~(\ref{asymE})], we treat it as having four times the
mass.  It is then a straightforward matter to calculate the fractional
energy transfer.  The result is compared with the actual data from
numerical simulations in Fig.~\ref{energyTransFig}. 
We see that the model captures the important features of
the collisions.  It accurately predicts that if the two shells collide
while traveling in the same direction, the faster one will lose energy.
It also agrees with the data in predicting that
in a head-on collision the SU(4) cloud will lose almost all of its
energy for large values of $E_a/E_\delta$.  Finally, it correctly
asserts that for a head-on collision there is a critical value of the
initial energy ratio below which the direction of energy transfer is
reversed.

This model works better than one might have hoped, but there is no
mystery as to why the predicted and observed energy transfer
disagree. First, the cloud trajectories are only approximate, and are
altered by additional interaction terms that only become significant
when the cloud radii are comparable. Second, the interactions are not
truly instantaneous, but occur over a finite time interval as the
clouds move through one another. Let us modify the statement of
conservation of energy of the clouds by including an inelastic term
$\Delta$, defined in terms of the initial and final cloud velocities
by
\begin{equation}
    \frac12 v_{\delta i}^2 + \frac12 v_{ai}^2 = 
    \frac12 v_{\delta f}^2 + \frac12 v_{af}^2  + \Delta  \, .
\end{equation}
Because all of the terms in the conservation of energy equation are
quadratic in velocities, we looked for an expression for $\Delta$
that was quadratic in the initial velocities and that provided good
agreement with the observed energy transfer. By trial and error, we
found that taking $\Delta = -{1\over 5} (3 v_{\delta i}^2 +
4v_{\delta i}v_{ai} -4 v_{ai}^2)$ provides excellent agreement with
the results obtained from numerical simulations, as can be seen from
the plots in Fig.~\ref{energyTransFig}.  The exact dynamics that give
rise to this this formula are still unclear to us.

We argued previously that the interaction between the clouds is
largely restricted to the time when $\delta - \tilde a$ was itself of
order $\tilde a$; this gives us a measure of the thickness of the
clouds.  To describe this more precisely let us define the beginning
of the interaction to be the time when 20\% of the total energy has
been transferred from one cloud to the other, and the end of the
interaction to be the time when 80\% has been transferred.  We also
define $\delta_0$ and $\tilde a_0$ to be the values of these variables
at the beginning of the interaction and 
\begin{equation}
    \rho = {\delta_0 -  \tilde a_0 \over \tilde a_0 } \, .
\end{equation}

\begin{figure}
\centering
\begin{tabular}{cc}
\epsfig{file= 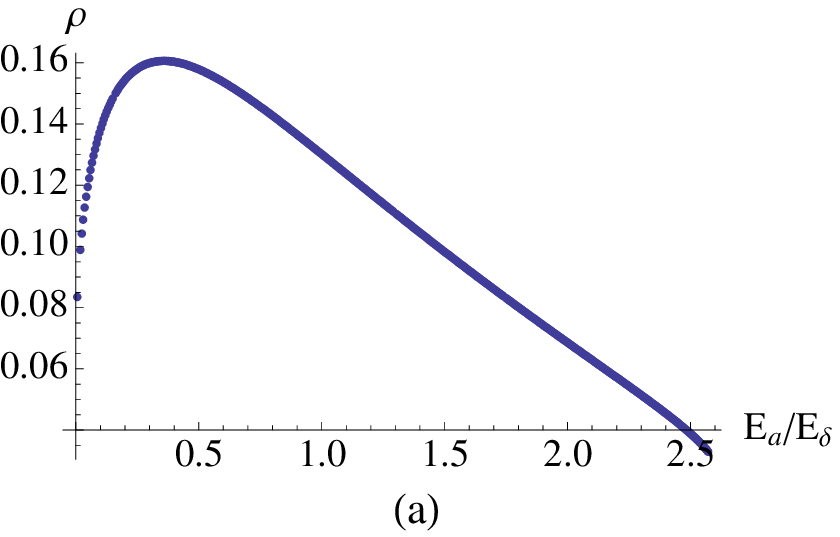,width=0.49\linewidth,clip=} & 
\epsfig{file= 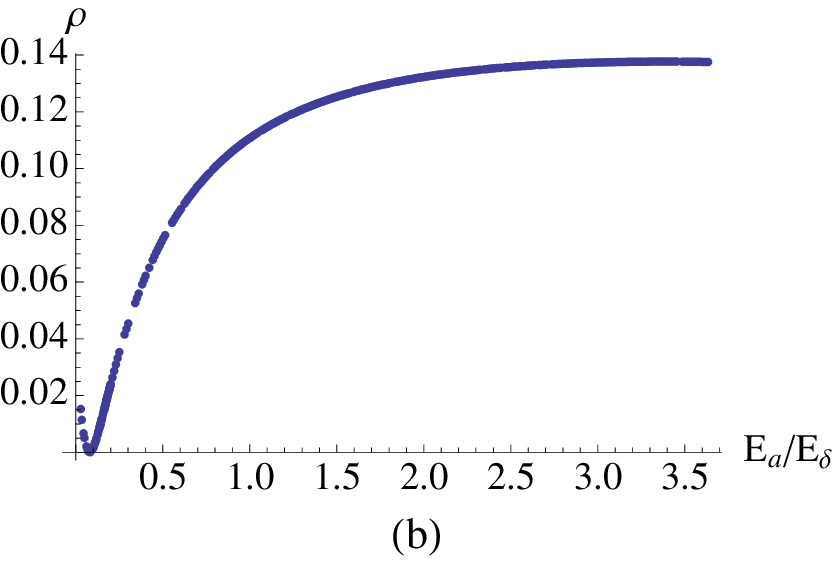,width=0.49\linewidth,clip=} 
\end{tabular}
\caption{The parameter $\rho$, which is a measure of cloud thickness, as a 
function of the energy ratio.  The result for two collapsing clouds is shown
in (a), and for a collapsing SU(4) cloud and expanding Dancer cloud in (b). }
\label{RhoFig}
\end{figure}

Figure~\ref{RhoFig} shows $\rho$ as a function of energy ratio for two
different regimes.  The left plot shows $\rho$ for an interaction in
which a collapsing SU(4) cloud overtakes a collapsing Dancer
cloud. Because the cloud velocities are equal when $E_a/E_\delta = 4$,
approaching this value from below corresponds to decreasing the
relative velocity of the clouds.  This plot therefore shows that
$\rho$ decreases as the relative velocity is decreased.  We see that
the two clouds can approach quite close to one another before
exchanging a significant amount of energy if they are moving slowly
relative to one another.  The right plot is for an interaction in
which a collapsing SU(4) cloud collides with an expanding Dancer
cloud.  In this case the relative velocity increases with
$E_a/E_\delta$.  We see that as this increases the clouds begin to
transfer their energy sooner, and hence at a greater separation.  The
maximum value of $\rho$ is about 0.17 in the former case and 0.15 in
the latter.  The similarity of the values for these two collision
scenarios would seem to indicate that the cloud thickness is a
relatively small fraction of the cloud radius, approximately
$(0.15-0.20) a_0$.

The behavior described by these two plots suggests that the clouds act
as dissipative media, and that when the SU(4) cloud moves through
the Dancer cloud, the energy loss increases with the relative velocity
of the clouds. This explains why when both clouds are collapsing
and their relative velocity is small, they can come very close
together before significant energy is transferred. In the other
situation, where the clouds collide head-on, the energy transfer
begins very quickly because the relative velocity is large.  This is
also consistent with the behavior of the inelasticity in the
collisions that we found previously.

\section{Summary and concluding remarks}
\label{conclude}

In this paper we have used moduli space methods to investigate the
properties of the massless magnetic monopoles that arise when a gauge
theory is spontaneously broken to a non-Abelian subgroup.  We have
shown how the natural metric on the Nahm data for a class of
SU($2M+2$) solutions with $2M$ massive and $M(2M-1)$ massless
monopoles can be obtained from the metric of a simpler class of
SU($M+1$) solutions.  Using this approach, we have explicitly verified
for the SU(4) (1,[1],1) case  
that the moduli spaces for the Nahm data and for the BPS solutions
are isomorphic, thus lending further
support to the conjecture that such an isomorphism holds in general.
We then applied this method to the problem of obtaining the metric for
the SU(6) (2,[2],[2],[2],2) solutions from the (2,[1]) SU(3) metric
studied by Dancer.  This gave us an effective Lagrangian for a class
of axially symmetric solutions.  This Lagrangian was then used to
study the interactions of the clouds that are the semiclassical
manifestation of the massless monopoles.

By examining explicit spacetime solutions, it has been known for some
time that the spacetime fields evaluated at the cloud radius are not
qualitatively different from those at points slightly further from or closer to
the origin.  One might therefore expect the interactions between
clouds to take place as if the shells of these clouds were
diffuse.  However, our simulations show instead that the clouds
interact more like relatively thin, hard shells.  In the collisions
between an SU(4) cloud and the Dancer clouds the energy transfer takes
place over a short interval before and after the cloud radii coincide,
suggesting an effective cloud thickness that is roughly 15-20\% of the
cloud radius.

Some intriguing open questions remain.  It is known that in Type IIB
string theory one can interpret D1-branes stretched between D3-branes
as the analogs of massive magnetic monopoles.  This suggests that
massless monopoles should, in some sense, correspond to D1-branes of
zero length connecting coincident D3-branes.  It would be desirable to
clarify these ideas, and to see if they would help explain the
properties of the clouds that we have found.  One would also
like to understand better the role of massless monopoles in the
electric-magnetic duality of $N = 4$ supersymmetric Yang-Mills theory,
where they should be the duals of the ``gluons'', the massless gauge
particles of the unbroken subgroup.  We hope that our results will
help shed light on these questions.

\begin{acknowledgments}

This work was supported in part by the U.S.~Department of Energy.

\end{acknowledgments}

\appendix*

\section{Calculation of $\bm{L_{\rm MS, eff}}$ for the SU(6) example}

In this appendix we present the details of the calculation of the
moduli space effective Lagrangian, Eq.~(\ref{LMSeffDef}), for the cylindrically
symmetric SU(6) solutions of Sec.~\ref{cylindrical}.

We begin by calculating the $C_{ab}$.  Because none of the $y^a$ tangent 
vectors require a compensating gauge action,  Eqs.~(\ref{metricform}) and
(\ref{jumpmetric}) for the metric reduce to
\begin{equation}
    C_{ab} =  4\pi\int_{s_L}^{s_0} ds \, \Tr {\partial T_\mu^L \over \partial y^a}
                 {\partial T_\mu^L \over \partial y^b}
            +  4\pi\int_{s_0}^{s_R} ds \, \Tr {\partial T_\mu^R \over \partial y^a}
                 {\partial T_\mu^R \over \partial y^b}
               + 2\pi \, \Tr \left({\partial A\over \partial y^a}
                        {\partial A^\dagger\over \partial y^b}
  + {\partial A\over \partial y^b} {\partial A^\dagger\over \partial y^a} \right) \, .
\label{su6metricform}
\end{equation}

The only nonvanishing contributions from the integral over the left interval are
\begin{equation}
    I_{ZZ}^L \equiv 4\pi \int_{s_0}^{s_L} ds \, 
         \Tr \left({\partial T_\mu^L \over \partial Z_L}\right)^2
\end{equation}
and 
\begin{equation}
    I_{DD}^L \equiv  4\pi \int_{s_0}^{s_L} ds \, 
     \Tr \left({\partial T_\mu^L \over \partial D_L}\right)^2  \, .
\end{equation}
(The mixed integral $I_{ZD}^L$ is zero because its integrand vanishes point
by point as a result of the trace.) 
Noting that 
\begin{equation}
    {\partial T^{DL}_\mu \over \partial Z_L} =
   \delta_{\mu 3} \, {\rm I}_2   \, ,
\end{equation}
we see immediately that 
\begin{equation}
   I_{ZZ}^L = 8\pi(s_0 -s_L) = 2 M_L \, .
\end{equation}

Equation~(\ref{axialT}) implies that 
\begin{equation}
    I_{DD}^L = 2\pi\int_{s_0}^{s_L} ds \, 
     \left[ 2\left({\partial g_1 \over \partial D_L}\right)^2 
       +  \left({\partial g_3 \over \partial D_L} \right)^2 \right]  \, .
\end{equation}
Defining $u = s - s_0$ and referring to Eq.~(\ref{axialfunctions}), we see that 
for the axially symmetric solutions
\begin{equation}
    {\partial g_j \over \partial D_L} = {1 \over D_L} 
        \left( g_j + u g_j' \right)  \, ,
\end{equation}
with the prime indicating differentiation with respect to $u$.  Hence,
\pagebreak
\begin{eqnarray}
    I_{DD}^L &=& {2\pi\over D_L^2} \int_0^{M_L/(4\pi)} du 
         \left[ (2 g_1^2  + g_3^2) + 2u (2 g_1 g_1' +  g_3 g_3') 
               +u^2(2 g_1^{'2} +  g_3^{'2}) \right]  \cr
          &=&{2\pi\over D_L^2} \int_0^{M_L/(4\pi)} du \left[ g_3^2 
      + {d \over du} \left(2ug_1^2 + u^2 g_1^2 g_3 \right) \right]   \, .
\end{eqnarray}
[In the second equality we have used Eq.~(\ref{topeq}) and its cyclic
permutations.]  This is now easily integrated to give
\begin{equation}
   I_{DD}^L = \cases{ \high
    {M_L\over 2}  (\mu_L - \sin \mu_L \cos \mu_L)(\tan \mu_L -\mu_L) 
       \left({\cos \mu_L \over \mu_L \sin^3 \mu_L} \right)  
       \, , & $\kappa_L = 0 \, , $  \cr\cr \high
    {M_L\over 2}  (\mu_L - \sinh \mu_L \cosh \mu_L)(\tanh \mu_L -\mu_L)
       \left({\cosh \mu_L \over \mu_L \sinh^3 \mu_L} \right)
       \, , & $\kappa_L = 1 \, , $  }
\label{IDDdef}
\end{equation}
where $\mu_L = M_L D_L/(4\pi)$.
   
The integrals on the right interval can be evaluated in the same
manner, and give the same result, except for the replacement of $M_L$,
$\mu_L$, and $\kappa_L$ by $M_R$, $\mu_R$, and $\kappa_R$,
respectively.

We need some limiting values of $I_{DD}$.  For $\kappa=0$ and $\mu$
close to $\pi$,
\begin{equation}
     I_{DD} = {\pi M\over 2(\pi - \mu)^3} 
         \left[ 1 + O\left({1 \over \pi -\mu} \right)\right]  \, ,
\label{IDDtriglimit}
\end{equation}
while for $\kappa=1$ and large $\mu$, 
\begin{equation}
     I_{DD} =  {M \over 2}\left[ 1 + O\left({1 \over \mu} \right)\right]  \, .
\label{IDDhyperlimit}
\end{equation}

To calculate the contribution from $A= K^{1/2}$, we recall from
Eq.~(\ref{Kdisplay}) that $K$ can be written in the block diagonal form
\begin{equation}
     K = \left( \matrix{ \lambda_1 & 0 & 0  \cr 
                  0 & \tilde K & 0 \cr
                  0 & 0 & \lambda_2 } \right)  \, ,
\end{equation}  
where the $2 \times 2$ matrix $\tilde K$ can be expanded in terms of Pauli 
matrices as 
\begin{equation} 
     \tilde K  = (p-C) {\rm I}_2 + 2B \rho_x + (R-q) \rho_z  \, .
\end{equation}
This can be rewritten as
\begin{equation}
     \tilde K = U^{-1} P  U  \, ,
\end{equation}
where $U = \exp{i \theta \rho_y /2}$ with $\tan \theta = 2B /(R-q)$ and 
$P$ is a diagonal matrix with eigenvalues
\begin{equation}
     \lambda_\pm  = (p-C) \pm \sqrt{4B^2 + (R-q)^2}  \, .
\label{lamPlusMin}
\end{equation}
The square root of $K$ is also block diagonal, with the middle block being
$\tilde K^{1/2} = U^{-1} P^{1/2} U$, whose
derivatives are
\begin{equation}
    \partial_a \tilde K^{1/2} = U^{-1} (\partial_a P^{1/2}) U
     + {i \over 2} (\partial_a \theta) [U^{-1} P^{1/2} U, \rho_y]  \, .
\label{KtildeDeriv}
\end{equation}
To calculate the metric we need 
\pagebreak
\begin{eqnarray} 
    \Tr \partial_a \tilde K^{1/2} \partial_b \tilde K^{1/2} 
        &=& \Tr \partial_a P^{1/2} \partial_b P^{1/2}  
           + {i \over 2} (\partial_a \theta) 
               \Tr \left( \partial_b P^{1/2} [P^{1/2}, \rho_y]\right) \cr 
        &&\quad + {i \over 2} (\partial_b \theta)
               \Tr \left( \partial_a P^{1/2} [P^{1/2}, \rho_y]\right) 
         - {1 \over 4}(\partial_a \theta)(\partial_b \theta)
           \Tr [P^{1/2}, \rho_y]^2   \, . \cr &&
\end{eqnarray}
Because both $P$ and $\partial_a P$ are diagonal, the middle two terms on 
the right-hand side both vanish.  The remaining terms give
\begin{equation} 
     \Tr \partial_a \tilde K^{1/2} \partial_b \tilde K^{1/2}
        = \partial_a \sqrt{\lambda_+} \, \partial_b \sqrt{\lambda_+}
        + \partial_a \sqrt{\lambda_-} \, \partial_b \sqrt{\lambda_-} 
        + {1 \over 2} (\partial_a \theta)(\partial_b \theta)
     \left( \sqrt{\lambda_+} - \sqrt{\lambda_-} \right)^2  \, .
\end{equation}
Adding to this the contributions from the corner elements
of $K$ gives
\begin{equation} 
     \Tr \partial_a A \,\partial_b A =
     \Tr \partial_a  K^{1/2} \partial_b  K^{1/2} 
       =  \sum_\sigma { \partial_a \lambda_\sigma \, \partial_b \lambda_\sigma
           \over 4\lambda_\sigma}
     + {1 \over 2} \partial_a \theta \, \partial_b \theta
     \left( \sqrt{\lambda_+} - \sqrt{\lambda_-} \right)^2 \, ,
\end{equation}
where the $\lambda_\sigma$ are the four eigenvalues of $K$.  

Combining this with our previous results, we obtain.
\begin{eqnarray}
   C_{ab} \, dy^a dy^b &=&   2 M_L\, dZ_L^2 + 2M_R \, dZ_R^2  + I_{DD}^L \, dD_L^2
     + I_{DD}^R \, dD_R^2  \cr\cr  \quad &+&
   \pi \left[\sum_\sigma  { \partial_a \lambda_\sigma \, \partial_b \lambda_\sigma
           \over \lambda_\sigma} + 
        2  \left( \sqrt{\lambda_+} - \sqrt{\lambda_-} \right)^2 
    \partial_a \theta \, \partial_b \theta \right] dy^a dy^b  \, .
\end{eqnarray}
      
The next step is to calculate the $B_{2a}$.  These only get a contribution from 
the boundary term, and are given by 
\begin{eqnarray}
    B_{2a} &=&  -{2\pi i}\, \Tr \left(
        \left[ {\partial A \over \partial y^a} , A \right] t_2 \right)  \cr
    &=&  -{2\pi i}\, \Tr \left(\left[ {\partial \tilde K^{1/2} 
             \over \partial y^a} , \tilde K^{1/2} \right] \rho_y \right)  \, .
\end{eqnarray}
With the aid of Eq.~(\ref{KtildeDeriv}), this can be rewritten as 
\begin{eqnarray}
    B_{2a} &=& \pi \, \partial_a \theta \,
     \Tr \left\{ \left[ \left[P^{1/2}, \rho_y \right], P^{1/2} \right] \rho_y 
            \right\}   \cr\cr
       &=& \pi \, \partial_a \theta \,
     \Tr \left[P^{1/2}, \rho_y \right]^2 \cr \cr 
     &=&  {2\pi \,} \partial_a \theta  \,
        \left( \sqrt{\lambda_+} - \sqrt{\lambda_-} \right)^2  \, .
\end{eqnarray}

We also need $E_{22}^{-1}$.  Referring to Eqs.~(\ref{Etwotwo}) and
(\ref{lamPlusMin}), we see that
\begin{equation}
    E_{22}^{-1} ={1 \over 4\pi( \lambda_+ + \lambda_-)}   \, .
\end{equation}

Using these last two results, we can calculate the correction term that
converts $L_{\rm MS}$, Eq.~(\ref{LMS}), to $L_{\rm MS,eff}$, Eq.~(\ref{LMSeffDef}).
The terms quadratic in $\partial_a \theta$ combine nicely, and we find that 
\pagebreak
\begin{eqnarray}
  \left[ C_{ab} - B_{a2}E_{22}^{-1} B_{2b} \right]\, dy^a dy^b
      &=&   2 M_L\, dZ_L^2 + 2M_R \, dZ_R^2  + I_{DD}^L \, dD_L^2
     + I_{DD}^R \, dD_R^2  \cr\cr  \quad &+&
   \pi \left[\sum_\sigma  { \partial_a \lambda_\sigma \, \partial_b \lambda_\sigma
           \over \lambda_\sigma} + 
     {\left(\lambda_+  - \lambda_- \right)^2 
           \over \left(\lambda_+ + \lambda_- \right)}
    \partial_a \theta \, \partial_b \theta \right] dy^a dy^b  \, .
\end{eqnarray}


\begin{thebibliography}{99}

\bibitem{Lee:1996vz}
  K.~Lee, E.~J.~Weinberg and P.~Yi,
  Phys.\ Rev.\  D {\bf 54}, 6351 (1996).

\bibitem{Weinberg:1979zt}
  E.~J.~Weinberg,
  Nucl.\ Phys.\  B {\bf 167}, 500 (1980).

\bibitem{Lu:1998br}
  C.~Lu,
  Phys.\ Rev.\  D {\bf 58}, 125010 (1998)

\bibitem{Houghton:1999qu}
  C.~Houghton, P.~W.~Irwin and A.~J.~Mountain,
  JHEP {\bf 9904}, 029 (1999)


\bibitem{Weinberg:1982jh}
  E.~J.~Weinberg,
  Phys.\ Lett.\  B {\bf 119}, 151 (1982).

\bibitem{Weinberg:1998hn}
  E.~J.~Weinberg and P.~Yi,
  Phys.\ Rev.\  D {\bf 58}, 046001 (1998).


\bibitem{Dancer:1992kj}
  A.~S.~Dancer,
  Nonlinearity {\bf 5}, 1355 (1992).

\bibitem{Dancer:1992kn}
  A.~S.~Dancer,
  Commun.\ Math.\ Phys.\  {\bf 158}, 545 (1993).

\bibitem{Dancer:1992hf}
  A.~S.~Dancer and R.~A.~Leese,
  Proc.\ Roy.\ Soc.\ Lond.\ A {\bf 440} (1993) 421.

\bibitem{Dancer:1997zx}
  A.~S.~Dancer and R.~A.~Leese,
  Phys.\ Lett.\  B {\bf 390}, 252 (1997).

\bibitem{Houghton:2002bz}
  C.~J.~Houghton and E.~J.~Weinberg,
  Phys.\ Rev.\  D {\bf 66}, 125002 (2002).

\bibitem{Manton:1981mp}
  N.~S.~Manton,
  Phys.\ Lett.\  B {\bf 110}, 54 (1982).

\bibitem{Chen:2001ge}
  X.~Chen and E.~J.~Weinberg,
  Phys.\ Rev.\  D {\bf 64}, 065010 (2001).

\bibitem{Manton:1988bn}
  N.~S.~Manton and T.~M.~Samols,
  Phys.\ Lett.\  B {\bf 215}, 559 (1988).

\bibitem{Stuart:1994tc}
  D.~Stuart,
  Commun.\ Math.\ Phys.\  {\bf 166}, 149 (1994).



\bibitem{Chen:2001qt}
  X.~Chen, H.~Guo and E.~J.~Weinberg,
  Phys.\ Rev.\  D {\bf 64}, 125004 (2001).


\bibitem{Atiyah:1985dv}
  M.~F.~Atiyah and N.~J.~Hitchin,
  Phys.\ Lett.\  A {\bf 107}, 21 (1985).

\bibitem{Lee:1996if}
  K.~Lee, E.~J.~Weinberg and P.~Yi,
  Phys.\ Lett.\  B {\bf 376}, 97 (1996).

\bibitem{Gauntlett:1996cw}
  J.~P.~Gauntlett and D.~A.~Lowe,
  Nucl.\ Phys.\  B {\bf 472}, 194 (1996).


\bibitem{Nahm:1982jb}
  W.~Nahm,
   ``The construction of all self-dual multimonopoles by the ADHM method'',
 in {\it Monopoles in Quantum Field Theory}, eds. N.~S.~Craigie et
al.   (World Scientific, Singapore, 1982).


\bibitem{Nahm:1981xg}
  W.~Nahm,
  ``Multimonopoles in the ADHM construction,''
in {\it Gauge Theories and Lepton Hadron Interactions},
eds.  Z. Horvath et al. (Central Research Institute for Physics, Budapest,
1982).


\bibitem{Nahm:1981nb}
  W.~Nahm,
  ``All self-dual multimonopoles for arbitrary gauge groups,''
in {\it Structural Elements in Particle Physics and Statistical
Mechanics}, eds. J. Honerkamp et al. (Plenum, New York, 1983).

\bibitem{Nahm:1983sv}
  W.~Nahm,
  ``Self-dual monopoles and calorons,''
in {\it Group Theoretical Methods in Physics}, eds. G.~Denardo et
 (Springer-Verlag, Berlin, 1984).

\bibitem{nakajima}
H.~Nakajima, ``Monopoles and Nahm's equations'', in {\it Einstein Metrics and
Yang-Mills Connections}, T.~Mabuchi and S.~Mukai eds. (Marcel Dekker, New York
1993).

\bibitem{takahasi}
   M. Takahasi, Ph.D Thesis, University of Tokyo.

\bibitem{Lee:1996kz}
  K.~Lee, E.~J.~Weinberg and P.~Yi,
  Phys.\ Rev.\  D {\bf 54}, 1633 (1996).



\bibitem{Weinberg:2006rq}
  E.~J.~Weinberg and P.~Yi,
  Phys.\ Rept.\  {\bf 438}, 65 (2007).


\bibitem{Irwin:1997ew}
   P.~Irwin,
  Phys.\ Rev.\ D {\bf 56}, 5200 (1997).


\end{thebibliography}
\end{document}